\documentclass[a4paper, 10pt, reqno, DIV=calc]{amsart}

\usepackage[utf8]{inputenc}
\usepackage[T1]{fontenc}
\usepackage{lmodern}
\usepackage{microtype}
\usepackage{amsmath, amsthm, amssymb, thmtools}
\usepackage{relsize}
\usepackage{enumitem}		
\usepackage[mathscr]{eucal}

\usepackage[dvipsnames]{xcolor}
\usepackage[linktocpage]{hyperref}
\definecolor{colorred}{HTML}{B00000}
\definecolor{colorgreen}{HTML}{258300}
\definecolor{colorblue}{HTML}{2e32fa}
\hypersetup{colorlinks=true, linkcolor=colorred, citecolor=colorgreen, urlcolor=black}								
\usepackage{xifthen}								
\usepackage{url}
\usepackage{todonotes}
\usepackage{caption}
\usepackage{bbm}									
\usepackage{stmaryrd}								
\usepackage{upgreek}

\makeatletter
\newcommand\MyAutoefPhrasecolorGroup[1]{%
  \color@begingroup\color{MyCurrentcolor}#1\endgroup
}%
\def\HyRef@testreftype#1.#2\\{%
 \colorlet{MyCurrentcolor}{.}%
 \ltx@IfUndefined{#1autorefname}{%
   \ltx@IfUndefined{#1name}{%
     \HyRef@StripStar#1\\*\\\@nil{#1}%
     \ltx@IfUndefined{\HyRef@name autorefname}{%
       \ltx@IfUndefined{\HyRef@name name}{%
         \def\HyRef@currentHtag{}%
         \Hy@Warning{No autoref name for `#1'}%
       }{%
         \edef\HyRef@currentHtag{%
           \noexpand\MyAutoefPhrasecolorGroup{%
             \expandafter\noexpand\csname\HyRef@name name\endcsname
           }%
           \noexpand~%
         }%
       }%
     }{%
       \edef\HyRef@currentHtag{%
         \noexpand\MyAutoefPhrasecolorGroup{%
           \expandafter\noexpand
           \csname\HyRef@name autorefname\endcsname
         }%
         \noexpand~%
       }%
     }%
   }{%
     \edef\HyRef@currentHtag{%
       \noexpand\MyAutoefPhrasecolorGroup{%
         \expandafter\noexpand\csname#1name\endcsname
       }%
       \noexpand~%
     }%
   }%
 }{%
   \edef\HyRef@currentHtag{%
     \noexpand\MyAutoefPhrasecolorGroup{%
       \expandafter\noexpand\csname#1autorefname\endcsname
     }%
     \noexpand~%
   }%
 }%
}%
\makeatother

\numberwithin{equation}{section}

\newcommand{\nlc}{{\ensuremath{\textnormal{c}}}}

\newcommand{\nlC}{{\ensuremath{\textnormal{C}}}}

\newcommand{\rmc}{{\ensuremath{\mathrm{c}}}}
\newcommand{\rmd}{{\ensuremath{\mathrm{d}}}}
\newcommand{\rme}{{\ensuremath{\mathrm{e}}}}

\newcommand{\rmD}{{\ensuremath{\mathrm{D}}}}
\newcommand{\rmE}{{\ensuremath{\mathrm{E}}}}

\newcommand{\sfd}{{\ensuremath{\mathsf{d}}}}
\newcommand{\sfe}{{\ensuremath{\mathsf{e}}}}

\newcommand{\sfr}{{\ensuremath{\mathsf{r}}}}

\newcommand{\sfB}{{\ensuremath{\mathsf{B}}}}

\newcommand{\scrD}{{\ensuremath{\mathscr{D}}}}

\newcommand{\scrK}{{\ensuremath{\mathscr{K}}}}

\newcommand{\scrM}{{\ensuremath{\mathscr{M}}}}

\newcommand{\scrP}{{\ensuremath{\mathscr{P}}}}

\newcommand{\scrS}{{\ensuremath{\mathscr{S}}}}
\newcommand{\scrT}{{\ensuremath{\mathscr{T}}}}

\newcommand{\scrX}{{\ensuremath{\mathscr{X}}}}

\newcommand{\bdpi}{{\ensuremath{\boldsymbol{\pi}}}}

\newcommand{\N}{\boldsymbol{\mathrm{N}}}						
\newcommand{\R}{\boldsymbol{\mathrm{R}}}						
\renewcommand{\d}{\,\mathrm{d}}				

\let\limsup\undefined
\let\liminf\undefined

\DeclareMathOperator*{\limsup}{limsup}		
\DeclareMathOperator*{\liminf}{liminf}		
\DeclareMathOperator{\supp}{spt}			

\theoremstyle{definition}
\newtheorem{bump}{Bump}[section]

\theoremstyle{plain}
\newtheorem{theorem}[bump]{Theorem}
\newtheorem{proposition}[bump]{Proposition}
\newtheorem{definition}[bump]{Definition}
\newtheorem{lemma}[bump]{Lemma}
\newtheorem{corollary}[bump]{Corollary}

\theoremstyle{remark}
\newtheorem{remark}[bump]{Remark}

\makeatletter
\def\nonumberfootnote{\xdef\@thefnmark{}\@footnotetext}			
\makeatother

\newcommand{\mms}{\mathit{M}}				
\newcommand{\met}{\sfd}						
\newcommand{\Rmet}{g}						
\newcommand{\meas}{\mathfrak{m}}				
\newcommand{\mmeas}{\mathfrak{n}}
\newcommand{\Leb}{\mathscr{L}}				
\newcommand{\vol}{\mathrm{vol}}				
\newcommand{\Prob}{\mathscr{P}}		        
\newcommand{\OptGeo}{\mathrm{OptGeo}}		
\newcommand{\OptTGeo}{\mathrm{OptTGeo}}
\newcommand{\Id}{\mathrm{Id}}				
\newcommand{\ac}{{\mathrm{ac}}}

\newcommand{\TCD}{\mathrm{TCD}}

\newcommand{\TMCP}{\mathrm{TMCP}}

\newcommand{\comp}{\nlc}					
\newcommand{\loc}{\mathrm{loc}}				
\newcommand{\pr}{\mathrm{proj}}				
\newcommand{\Ric}{\mathrm{Ric}}				
\newcommand{\Cont}{\nlC}					
\newcommand{\Ell}{\mathit{L}}				
\newcommand{\Geo}{\mathrm{Geo}}				

\DeclareMathOperator{\Hess}{Hess}			
\newcommand{\eval}{\sfe}					
\newcommand{\push}{\sharp}					

\newcommand{\Len}{\mathrm{Len}}
\newcommand{\TGeo}{\mathrm{TGeo}}
\newcommand{\tsep}{\uptau}

\usepackage{blindtext}

\allowdisplaybreaks

\providecommand{\bysame}{\leavevmode\hbox to3em{\hrulefill}\thinspace}

\let\oldtocsection=\tocsection

\let\oldtocsubsection=\tocsubsection

\let\oldtocsubsubsection=\tocsubsubsection

\renewcommand{\tocsection}[2]{\hspace{0em}\oldtocsection{#1}{#2}}
\renewcommand{\tocsubsection}[2]{\hspace{1em}\oldtocsubsection{#1}{#2}}
\renewcommand{\tocsubsubsection}[2]{\hspace{2em}\oldtocsubsubsection{#1}{#2}}

\newcommand{\nocontentsline}[3]{}
\newcommand{\tocless}[2]{\bgroup\let\addcontentsline=\nocontentsline#1{#2}\egroup}

\newcommand{\mres}{\mathbin{\vrule height 1.6ex depth 0pt width
0.13ex\vrule height 0.13ex depth 0pt width 1.3ex}}

\DeclareFontFamily{U}{mathb}{\hyphenchar\font45}
\DeclareFontShape{U}{mathb}{m}{n}{
<-6> mathb5 <6-7> mathb6 <7-8> mathb7
<8-9> mathb8  <9-11> mathb9
<11-12> mathb10 <12-> mathb12
}{}
\DeclareSymbolFont{mathb}{U}{mathb}{m}{n}
\DeclareMathSymbol{\llcurly}{\mathrel}{mathb}{"CE}
\DeclareMathSymbol{\ggcurly}{\mathrel}{mathb}{"CF}

\newcommand{\NEW}[2]{{#2}}

\begin{document}

\title[Timelike Ricci bounds for low regularity spacetimes]{Timelike Ricci bounds for low regularity spacetimes by optimal transport}
\author{Mathias Braun}
\address{Department of Mathematics, University of Toronto, 40 St. George Street Room 6290, Toronto, Ontario M5S 2E4, Canada}
\email{braun@math.toronto.edu}
\author{Matteo Calisti}
\address{Universität Wien, Institut für Mathematik,  Oskar-Morgenstern-Platz 1, 1090 Wien, Austria}
\email{matteo.calisti@univie.ac.at}
\date{\today}
\thanks{The authors are grateful to Clemens Sämann for various helpful comments on an earlier version of the paper.  M.B.~acknowledges funding by the Fields Institute for Research in Mathematical Sciences. His
research is supported in part by Canada Research Chairs Program funds and a Natural Sciences 
and Engineering Research Council of Canada Grant (2020-04162) held by Robert McCann. M.C. acknowledges the support of Project P 33594 of the Austrian Science Fund FWF}

\begin{abstract} We prove that a globally hyperbolic smooth spacetime endowed with a $\smash{\mathrm{C}^1}$-Lorentzian metric whose Ricci tensor is bounded from below in all timelike directions, in a distributional sense, obeys the timelike measure-contraction property. This result includes a class of spacetimes with borderline regularity for which local existence results for the vacuum Einstein equation are known in the setting of spaces with timelike Ricci bounds in a synthetic sense. In particular, these spacetimes satisfy timelike  Brunn--Min\-kowski, Bonnet--Myers, and Bishop--Gromov inequalities in sharp form, without any time\-like nonbranching assumption. 

If the metric is even $\smash{\mathrm{C}^{1,1}}$, in fact the stronger timelike curvature-dimension condition holds. In this regularity, we also obtain uniqueness of chronological optimal couplings and chronological geodesics.\end{abstract}

\maketitle
\thispagestyle{empty}

\tableofcontents

\addtocontents{toc}{\protect\setcounter{tocdepth}{1}}

\section{Introduction}\label{Ch:Intro} 

\subsection*{Background} In the last two decades, optimal transport theory has been applied to a large variety of mathematical areas, including PDEs, Riemannian geometry, numerical analysis, etc. More recently, it has revealed promising links to general relativity, i.e.~Einstein's theory of gravity, as follows. Let $\scrM$ be a smooth spacetime of dimension $n\in \N_{\geq 2}$, endowed with a globally hyperbolic Lorentzian metric $\Rmet$ 
--- physically, one should always think of $(\scrM,\Rmet)$ to solve the Einstein equation with given cosmological constant $\Lambda\in \R$ and energy-momentum tensor $T$. If $\Rmet$ is smooth\footnote{In fact, $\smash{\Cont^2}$-regularity suffices for the arguments in \cite{braun2022, mccann2020, mondinosuhr2018}.},  
\cite{mccann2020,mondinosuhr2018} and later \cite{braun2022, braunohta} showed convexity properties of certain entropy functionals with respect to the volume measure $\textnormal{vol}_\Rmet$ along ``chronologi\-cal'' geodesics in $\scrP(\scrM)$ to characterize the condition
\begin{align}\label{Eq:RK}
\Ric_\Rmet\geq K\textnormal{ in all timelike directions}.
\end{align}
Here, $\scrP(\scrM)$ is the space of Borel probability measures on $\scrM$. The relevant geometry thereon is described by a certain Lorentzian transport cost $\smash{\ell_{\Rmet,p}}$, where $p\in (0,1)$, which plays the role of a time separation function.

The condition \eqref{Eq:RK} has high relevance in general relativity. Indeed, for $\Lambda=0$, \eqref{Eq:RK} for $K=0$ is equivalent to the \emph{strong energy condition} of Hawking and Penrose \cite{hawking1973,hawking1970}. Moreover   \cite{cavalletti2020}, for arbitrary $\Lambda\in \R$, if $\inf\,\textnormal{scal}_\Rmet(\scrM) > -\infty$ then \eqref{Eq:RK} for $\smash{K =\inf \textnormal{scal}_\Rmet(\scrM)/2 - \Lambda}$ is implied by the \emph{weak energy condition} ``$T\geq 0$ in all timelike directions''. The latter is believed to hold for most physically reasonable $T$ \cite[p.~218]{wald1984}, and clearly holds in the vacuum case $T=0$. See \cite{carroll2004,hawking1973,mccann2020, mondinosuhr2018,treude2013,wald1984} for further discussions about \eqref{Eq:RK}.

On the other hand, said groundbreaking discoveries of \cite{mccann2020, mondinosuhr2018} have lead to the synthetic theory of TCD and TMCP spaces 
via the Boltzmann entropy  \cite{cavalletti2020} and later via the Rényi entropy \cite{braun2022} in the framework of measured Lorentzian spaces \cite{cavalletti2020, kunzinger2018}, i.e.~natural generalizations of spacetimes. The definitions in \cite{braun2022, cavalletti2020} are partly equivalent \cite[Thm.~3.35, Thm.~4.20]{braun2022}, yet the approach in \cite{braun2022} yields quantitatively stronger geometric properties \emph{a priori}, as made precise below. TCD and TMCP spaces are Lorentzian analogues of CD and MCP metric measure spaces \cite{erbar2015,lott2009, ohta2007, sturm2006a, sturm2006b}. Roughly speaking, to make sense the respective formulations of TCD and TMCP only require a space with an abstract notion of time separation function, encoding how geodesics of points on $\scrM$ and probability measures look like, and a reference measure, encoding how to ``average sectional curvatures''. Hence, these conditions are suitable to give a meaning to \eqref{Eq:RK} even when no smooth structure is available to define the inherent Ricci tensor itself.

\subsection*{Objective} The aim of this work is to provide a  first link of this novel synthetic point of view to the more customary analytic approach to \eqref{Eq:RK} for Lorentzian metrics of low regularity by distribution theory; see e.g.~\cite{graf2020, grosser2001, lefloch2007,steinbauer2008, stv} for previous works on distributional energy conditions. 

We will start with a Lorentzian metric $\Rmet$ on $\scrM$  of regularity at least $\smash{\Cont^1}$ obeying \eqref{Eq:RK} --- and weighted versions thereof --- in a distributional sense. Then we prove that the measured Lorentzian space canonically induced by $(\scrM,\Rmet)$ according to \autoref{Sec:Construction} below has timelike (Bakry--Émery--)Ricci curvature bounded from below by $K$ in the  indicated synthetic senses. 
As applications, inter alia we derive timelike geometric inequalities. Notably, these are obtained in \emph{sharp} form even though the regularity of $\Rmet$ might be below $\smash{\Cont^{1,1}}$, where $\Rmet$ might admit timelike branching \cite[Def.~1.10]{cavalletti2020}, and the localization technique \cite[Ch.~4, Sec.~5.3, Sec.~5.4]{cavalletti2020} used in \cite{cavalletti2020} to prove these sharp inequalities in the synthetic setting does not apply. 

This partly answers a question raised in \cite{kunzinger2022}. There \cite[Thm.~5.4]{kunzinger2022}, smooth manifolds with $\smash{\Cont^1}$-\emph{Riemannian} metrics and distributional Ricci bounds are shown to be CD spaces. A partial converse holds as well \cite[Thm.~6.3]{kunzinger2022}, yet a Lorentzian analogue of this is beyond the scope of our work.

Our main results provide a set of concrete examples of TCD and TMCP spaces beyond ``sufficiently regular'' spacetimes. Moreover, as concretized further below, the proofs of our main results   --- \autoref{Th:TMCP} and \autoref{Th:TCD nonbr} ---  are based on an approximation argument and are, as such,    heavily inspired by the proof of stability of TCD and TMCP  \cite{cavalletti2020}, see also \cite{braun2022}, under \NEW{the first time} the novel notion of weak convergence of measured Lorentzian spaces introduced in  \cite[Thm.~3.12]{cavalletti2020}; see \autoref{Re:Stab} for a discussion of how this relates to (open) stability questions. 

The mathematical relevance of our setting comes from the PDE point of view, where standard local existence results for the vacuum Einstein equation, together with Sobolev's embedding theorem, in four dimensions just grant $\smash{\Cont^1}$-regularity of $\Rmet$ \cite[p.~10]{rendall2005}, see also \cite{christodoulou2009,klainerman}. In general, since the Einstein equation is hyperbolic, its solutions are typically not smooth, which makes the synthetic TCD and TMCP framework  interesting to study its rough solutions. From a geometric perspective, $\smash{\Cont^1}$ \cite{graf2020} and $\smash{\Cont^{1,1}}$ \cite{kunzinger2015-1} are the lowest regularities under which the classical Hawking singularity theorem \cite{hawking1966,hawking1973} has been proven under distributional timelike Ricci bounds. (See \cite{graf2020,kunzinger2022b} for $\smash{\Cont^1}$-versions of the Hawking--Penrose singularity theorem, and \cite{sbr} for an  overview over singularity theorems in general relativity.) Incidentally, our results build a first bridge between  \cite{graf2020, kunzinger2015-1} and the synthetic Hawking singularity theorem for timelike nonbranching low regularity  spacetimes from \cite[Thm.~5.6, Cor.~5.13]{cavalletti2020}. Indeed, by \autoref{Th:TCD nonbr} and \autoref{Re:Geod equ}, the distributional $\smash{\Cont^{1,1}}$-versions from \cite{kunzinger2015-1} are included in \cite{cavalletti2020} (in the sense that the assumptions in \cite{cavalletti2020} really extend those of \cite{kunzinger2015-1}). In a similar kind, by \autoref{Th:TMCP},  timelike nonbranching $\smash{\Cont^1}$-spacetimes with distributional timelike Ricci bounds as in \cite{graf2020} are covered by \cite{cavalletti2020}. However, unlike the $\smash{\Cont^{1,1}}$-case,  $\smash{\Cont^1}$-spacetimes are generally expected to admit  timelike branching, hence \cite{cavalletti2020} remains unknown to apply to some spaces from \cite{graf2020}.


\subsection*{Results} Now we outline our main results. In order to keep the presentation light, we postpone technicalities and more precise definitions to \autoref{Ch:SPT low reg}.

Let $\Rmet$ be a globally hyperbolic Lorentzian metric on $\scrM$ with regularity at least $\smash{\Cont^1}$. We write $\smash{\leq_\Rmet}$ and  $\smash{\ll_\Rmet}$ for the future-directed $\Rmet$-causality and future-directed $\Rmet$-chronology on $\scrM$, respectively. Let $\tsep_\Rmet$ denote the usual time separation function induced by $\Rmet$, i.e.~if $x,y\in\mms$ obey  $\smash{x\leq_\Rmet y}$ then $\smash{\tsep_\Rmet(x,y)}$ constitutes the maximal $\Rmet$-length of all future-directed $\Rmet$-causal curves from $x$ to $y$.  Let $V\in\Cont^1(\scrM)$ and $N \in [n,\infty)$, and define the \emph{$N$-Bakry--\smash{Émery}--Ricci tensor} by
\begin{align*}
\Ric_{\Rmet}^{N,V} := \Ric_{\Rmet} + \Hess_{\Rmet} V + \frac{1}{N-n}\,\rmD V\otimes \rmD V.
\end{align*}
It is understood in a distributional sense; lower bounds on it à la \eqref{Eq:RK} are then formulated by requiring  $\smash{\langle \Ric_\Rmet^{N,V}(X,X),\mu\rangle} \geq \smash{K\,\langle\Rmet(X,X),\mu\rangle}$ for every smooth $\Rmet$-timelike vector field $X$ on $\scrM$ and every nonnegative, compactly supported volume density $\mu$. Of course, if $\Rmet$ is of class $\smash{\Cont^2}$, all expressions and inequalities make sense in the ordinary pointwise way.

To further ease the presentation, in the subsequent outline we restrict ourselves to the case $K=0$, $V=0$, and $N=n$. It is already interesting since it represents the strong energy condition of Hawking and Penrose in low regularity, because $\smash{\Ric_\Rmet^{N,V}}$ then simply becomes the usual Ricci tensor $\smash{\Ric_\Rmet}$. Our results --- with appropriate modifications --- will hold and be shown in all generality.

Under these simplifications, the formulation of our results still requires some further notation. For $p\in (0,1)$, let $\smash{\ell_{\Rmet,p}\colon\scrP(\scrM)^2\to [0,\infty]\cup\{-\infty\}}$ be the so-called \emph{$p$-Lorentz--Wasserstein distance} \cite{eckstein2017} defined through
\begin{align*}
\ell_{\Rmet,p}(\mu,\nu) := \sup  \big\Vert \tsep_\Rmet\big\Vert_{\Ell^p(\scrM^2,\pi)}.
\end{align*}
Here, the supremum is taken over all \emph{$\Rmet$-causal couplings} $\pi$ of $\mu$ and $\nu$. The latter notion means that $\pi$ is a Borel probability measure on $\smash{\scrM^2}$ with marginals $\mu$ and $\nu$ such that $\smash{x\leq_\Rmet y}$ for $\pi$-a.e.~$\smash{(x,y)\in\scrM^2}$. The role of $\smash{\ell_{\Rmet,p}}$ is to ``lift'' the $\Rmet$-causal structure of $\scrM$ to the space of mass distributions on $\scrM$. 
And it enables us to define \emph{timelike proper-time parametrized $\smash{\ell_{\Rmet,p}}$-geodesics} which, roughly speaking, are curves $(\mu_t)_{t\in[0,1]}$ in $\scrP(\scrM)$ such that for every $s,t\in[0,1]$ with $s<t$,
\begin{align*}
\ell_{\Rmet,p}(\mu_s,\mu_t) = (t-s)\,\ell_{\Rmet,p}(\mu_0,\mu_1) > 0.
\end{align*}
Lastly, the canonical volume measure $\smash{\vol_\Rmet}$ induced by $\Rmet$ is invoked through the \emph{$n$-Rényi entropy} $\smash{\scrS_\Rmet^n\colon \scrP(\scrM)\to [-\infty,0]}$; subject to the Lebesgue decomposition $\mu = \smash{\rho\,\vol_\Rmet + \mu^\perp}$, the latter is defined by
\begin{align*}
\scrS_\Rmet^n(\mu) := -\int_{\scrM}\rho^{-1/N}\d\mu =  -\int_{\scrM} \rho^{1-1/N}\d\vol_\Rmet.
\end{align*} 

\begin{theorem}\label{Th:TheMainTechnicalResult} Assume $\smash{\Rmet}$ to be  $\smash{\Cont^1}$, and suppose it  satisfies the distri\-butional strong energy condition   \eqref{Eq:RK}. Then for every $p\in(0,1)$ and every compactly supported, $\meas$-absolutely continuous mass distributions $\smash{\mu_0,\mu_1\in\scrP(\scrM)}$ such that $\smash{x\ll_\Rmet y}$ for every $x\in\supp\mu_0$ and every $y\in\supp\mu_1$, there exists a timelike proper-time parametrized $\smash{\ell_{\Rmet,p}}$-geodesic $(\mu_t)_{t\in[0,1]}$ connecting $\mu_0$ to $\mu_1$ such that every $t\in[0,1]$ satisfies
\begin{align*}
\scrS_\Rmet^n(\mu_t) \leq (1-t)\,\scrS_\Rmet^n(\mu_0) + t\,\scrS_\Rmet^n(\mu_1).
\end{align*}
\end{theorem}

In other words, the distributional strong energy condition for $\Rmet$ implies displacement convexity \cite{mccann} of the $n$-Rényi entropy with respect to $\smash{\vol_\Rmet}$ along appropriate ``timelike'' geodesics in $\scrP(\scrM)$.

\autoref{Th:TheMainTechnicalResult} is our main technical ingredient. It immediately implies our main results. First, the measured Lorentzian space induced by $(\scrM,\Rmet,\vol_\Rmet)$ satisfies the $p$-independent timelike measure-contraction property $\smash{\TMCP(0,n)}$ under the same assumptions on $\Rmet$ as in \autoref{Th:TheMainTechnicalResult}, cf.~\autoref{Th:TMCP}. Second, if $\smash{\Rmet}$ is $\smash{\Cont^{1,1}}$, we even get the stronger timelike curvature-dimension condition $\smash{\TCD_p(0,n)}$ in \autoref{Th:TCD nonbr}. As indicated above, both assert the displacement convexity of $\smash{\scrS_\Rmet^n}$ along ``timelike'' geodesics in $\scrP(\scrM)$ which connect an $\meas$-absolutely continuous $\mu_0$ to a Dirac mass $\mu_1$ ($\TMCP$), or two $\meas$-absolutely continuous $\mu_0$ and $\mu_1$ ($\TCD$), respectively, obeying  a chronology relation called \emph{$\Rmet$-timelike $p$-dualizability} \cite{cavalletti2020}, cf.~\autoref{Def:TL p dual}. In general, the latter is strictly weaker than the chronology hypothesis on $\mu_0$ and $\mu_1$ in \autoref{Th:TheMainTechnicalResult}, which forces all geodesics starting in $\supp\mu_0$ and ending in $\supp\mu_1$ to stay away from the $\Rmet$-light cone in a ``uniform'' manner.


\begin{remark}\label{Re:RedEnt} To be precise,  three different  timelike curvature-dimension conditions have been proposed in \cite{braun2022, cavalletti2020} for general $K\in\R$ and $N\in [1,\infty)$. While \cite{cavalletti2020} sets up the so-called  entropic timelike curvature-dimension condition $\smash{\TCD_p^e(K,N)}$ by \emph{concavity} properties of an exponentiated Boltzmann entropy similar to \cite{erbar2015}, \cite{braun2022} defines the timelike curvature-dimension condition $\smash{\TCD_p(K,N)}$ and its reduced version $\smash{\TCD_p^*(K,N)}$ after \cite{bacher2010, sturm2006b}. We refer the reader to \autoref{Re:Relations} for general (and expected) relations between these notions.

We will concentrate on showing the second named version in our work. Our proofs can easily be adapted to derive the other conditions as well. 

An analogous note applies  to the three timelike measure-contraction properties $\smash{\TMCP^e(K,N)}$, $\smash{\TMCP(K,N)}$, and $\smash{\TMCP^*(K,N)}$ from \cite{braun2022, cavalletti2020}.


\end{remark}

From this, we directly infer the subsequent timelike geometric inequalities under the assumption on $\Rmet$ from \autoref{Th:TheMainTechnicalResult}. 
They are stated in a precise and slightly more general form in \autoref{Cor:Sharp1}, \autoref{Cor:Sharp2}, and \autoref{Cor:Sharp3}.
\begin{itemize}
\item \textbf{Brunn--Minkowski inequality.} Let $\smash{A_0,A_1\subset\scrM}$ be two relatively compact Borel sets obeying $\smash{x\ll_\Rmet y}$ for every $\smash{x\in \bar{A}_0}$ and every $\smash{y\in \bar{A}_1}$. Given any $t\in[0,1]$, the set $A_t \subset\scrM$ of all $t$-intermediate points of future directed $\Rmet$-timelike geodesics starting in  $A_0$ and ending in $A_1$ obeys
\begin{align*}
\vol_\Rmet[A_t]^{1/n} \geq (1-t)\,\vol_\Rmet[A_0]^{1/n} + t\,\vol_\Rmet[A_1]^{1/n}.
\end{align*} 
\item \textbf{Bishop--Gromov inequality.} Let $x\in\scrM$, and let $E\subset\scrM$ be a compact set which is $\smash{\tsep_\Rmet}$-star shaped with respect to $x$. For $r>0$, let $v_r$ denote the measure of the truncated hyperboloid $\{y\in \scrM : x \leq_\Rmet y,\, \tsep_\Rmet(x,y) \leq r\} \cap E$ with respect to $\smash{\vol_\Rmet}$. Then the quantity $v_r/r^n$ is nonincreasing in $r>0$. In other words, for every $r,R > 0$ with $R > r$,
\begin{align*}
\frac{v_r}{v_R} \geq \frac{r^n}{R^n}.
\end{align*}
\end{itemize}
Assuming the more restrictive distributional energy condition $\smash{\Ric_\Rmet \geq K}$ in all time\-like directions for $K>0$, we also obtain the 
\begin{itemize}
\item \textbf{Bonnet--Myers inequality.} We have
\begin{align*}
\sup \tsep_\Rmet(\scrM^2) \leq \pi\sqrt{\frac{n-1}{K}}.
\end{align*}
\end{itemize}
In our low regularity framework, starting from a distributional energy condition à la \eqref{Eq:RK} the only comparable result we are aware of is a Bishop--Gromov inequality in the $\smash{\Cont^{1,1}}$-case \cite{GRAF}. On the other hand, in the synthetic approach to timelike Ricci curvature lower bounds all these estimates are standard consequences, as shown in \cite{cavalletti2020} and later in \cite{braun2022}. By virtue of the approach from \cite{braun2022} we follow as outlined in \autoref{Re:RedEnt} above, all three named inequalities are obtained in sharp form; cf. \autoref{Re:Sharp} for further details.

Lastly, if $\Rmet$ is $\smash{\Cont^{1,1}}$, in \autoref{Cor:UNIQNNN} we obtain uniqueness of chronological $\smash{\ell_{\Rmet,p}}$-optimal couplings, uniqueness of ``timelike'' geodesics in $\scrP(\scrM)$, and solvability of a Lorentzian Monge problem under distributional timelike (Bakry--Émery--)Ricci curvature lower bounds. Here, as already outlined above, the crucial feature of working with a $\smash{\Cont^{1,1}}$-metric is that it ensures timelike nonbranching, a natural condition to get said uniquenesses. In analogy to the Riemannian fact \cite[Thm. 2.3]{kunzinger2022}, these uniqueness claims should be true without curvature assumptions, yet we do not address this generalization in our work. An advantage of the curvature hypothesis plus timelike nonbranching is that the TCD inequality automatically holds \emph{pathwise} along ``timelike'' geodesics of probability measures \cite[Thm.~3.41]{braun2022}. 


\subsection*{Outline of the proof of \autoref{Th:TheMainTechnicalResult}} The argument for our main results relies on a suitable approximation of $\Rmet$ by \emph{smooth} Lorentzian metrics which locally do not violate \eqref{Eq:RK} too much. A very rough version of the relevant  \autoref{Le:Approx}, yet conveying the key ideas for now, is the following.

\begin{lemma}\label{Le:Handwaving} Assume $\Rmet$ to satisfy \eqref{Eq:RK}. Then there exist smooth Lorentzian metrics $\{\check{\Rmet}_\varepsilon : \varepsilon > 0\}$ such that $\check{\Rmet}_\varepsilon \to \Rmet$ in $\smash{\Cont_\loc^1(\scrM)}$ as $\varepsilon \to 0$ and with the following property. For every compact $C\subset\scrM$ and every $\delta,\kappa >0$, there exists $\varepsilon_0 > 0$ such that for every $\varepsilon \in (0,\varepsilon_0)$ and every $\smash{v\in T\scrM\big\vert_C}$, we have 
\begin{align}\label{Eq:INRTq}
\vert v\vert_{\check{\Rmet}_\varepsilon} \geq \sqrt{\kappa}\quad\Longrightarrow\quad \Ric_{\check{\Rmet}_\varepsilon}(v,v) \geq -\delta\,\big\vert v\big\vert_{\check{\Rmet}_\varepsilon}^2.
\end{align}
\end{lemma}

This approximation result itself, at least in the unweighted case, is not new. In \cite{graf2020,kunzinger2015-1}, it has been employed to prove Hawking's singularity theorem in $\smash{\Cont^1}$- and $\smash{\Cont^{1,1}}$-regularity, respectively. It is the technical reason for our imposed regularity on $\Rmet$; a version of it e.g.~for Lipschitz metrics remains unknown. The mentioned Riemannian result in $\smash{\Cont^1}$-regularity \cite[Thm.~5.4]{kunzinger2022} has been derived from a similar approximation procedure \cite[Thm.~4.3]{kunzinger2022}.



Our argument for \autoref{Th:TheMainTechnicalResult} follows the proof of \cite[Thm.~3.29]{braun2022} for the weak stability of the $\TCD$ condition. In \autoref{Sec:Recovery} below, given $\{\check{\Rmet}_\varepsilon : \varepsilon > 0\}$ as in \autoref{Le:Handwaving}, for $\mu_0$ and $\mu_1$ as hypothesized with 
\begin{align*}
\kappa \propto \inf\tsep_\Rmet(\supp\mu_0\times\supp\mu_1)>0,
\end{align*}
we construct a recovery family $\smash{\{(\mu_0^\varepsilon,\mu_1^\varepsilon) : \varepsilon > 0\}}$ of $\smash{\check{\Rmet}_\varepsilon}$-time\-like $p$-dualizable pairs $\smash{(\mu_0^\varepsilon,\mu_1^\varepsilon)}$ for $(\mu_0,\mu_1)$, where $\smash{\mu_0^\varepsilon}$ and $\smash{\mu_1^\varepsilon}$ are absolutely continuous with respect to the volume measure $\smash{\vol_{\check{\Rmet}_\varepsilon}}$. This is done in such a way that the unique \cite{mccann2020} $\smash{\ell_{\check{\Rmet}_\varepsilon,p}}$-optimal transport from $\smash{\mu_0^\varepsilon}$ to $\smash{\mu_1^\varepsilon}$ only matches points with $\smash{\tsep_{\check{\Rmet}_\varepsilon}}$-distance larger than $\kappa$. This property, combined with \autoref{Le:Handwaving}, ensures displacement semiconvexity --- with ``$\smash{\check{\Rmet}_\varepsilon}$-timelike Ricci lower bound'' $-\delta$ ---  of the $n$-Rényi  entropy with respect to $\smash{\vol_{\check{\Rmet}_\varepsilon}}$ between $\smash{\mu_0^\varepsilon}$ and $\smash{\mu_1^\varepsilon}$ for sufficiently small $\varepsilon >0$ (depending on the values of  $\delta$ and $\kappa$), which we establish by hand in \autoref{Sec:Displacement} following the argument for  \cite[Prop.~A.3]{braun2022}. Up to subsequences, it then remains to first let $\varepsilon\to 0$ and then $\delta \to 0$; the relevant inequalities are stable under these limits essentially because as a statement about convexity, they are of zeroth order  nature. In \autoref{Sec:Concl} we then conclude the desired \autoref{Th:TMCP} and \autoref{Th:TCD nonbr}.

\begin{remark}\label{Re:Stab} Despite the similarity of the outlined  argument with \cite[Thm.~3.29]{braun2022}, we stress that the measured Lorentzian space 
induced by $\smash{(\scrM,\check{\Rmet}_\varepsilon,\vol_{\check{\Rmet}_\varepsilon})}$, for \emph{fixed} $\varepsilon > 0$, is unclear to obey a TCD or TMCP condition, with lower bound $-\delta$ or otherwise.  Indeed, among others the possible range of $\varepsilon$ in \autoref{Le:Handwaving} depends on the parameter $\kappa$, which describes how far away mass distributions have to lie from each other in order for the timelike Ricci bound \eqref{Eq:INRTq} to be satisfied along their optimal transport. In particular, \autoref{Th:TheMainTechnicalResult} does \emph{not} follow from weak stability of the TCD condition (only from a similar proof). Hence, $\smash{\Cont^1}$-spacetimes are still unclear to fall into the class of ``timelike Ricci limit spaces'' after the convergence of \cite[Thm.~3.12]{cavalletti2020}, whose structure thus remains completely unstudied.
\end{remark}

\subsection*{Organization} In \autoref{Ch:SPT low reg}, we review basic notions of $\smash{\Cont^1}$-Lorentzian spacetimes and their Lorentzian geodesic structure, recall and slightly extend the approximation results from \cite{graf2020, kunzinger2015-1, kunzinger2022}, and outline basics of Lorentzian optimal transport. \autoref{Ch:Proofs} contains the proofs of \autoref{Th:TMCP}, \autoref{Th:TCD nonbr}, and consequences of these main results.

\addtocontents{toc}{\protect\setcounter{tocdepth}{2}}
\section{Spacetimes of low regularity}\label{Ch:SPT low reg}

\subsection{Terminology}\label{Sec:Terminology} By convention, all Lorentzian metrics in this paper will have signature $+,-,\dots,-$.

By $\scrM$ we denote a topological manifold (connected, Hausdorff, second countable) of class $\smash{\Cont^\infty}$. The latter is no loss of generality, since for generic vector fields to be continuous, $\smash{\Cont^1}$-regularity would be a natural assumption on $\scrM$, yet any $\smash{\Cont^1}$-manifold possesses a unique $\smash{\Cont^\infty}$-structure that is $\smash{\Cont^1}$-compatible with the given $\smash{\Cont^1}$-structure \cite[Thm.~2.9]{hirsch76}, and we would then simply work with that smooth atlas. 

All over this chapter, let $\Rmet$ be a Lorentzian metric on $\scrM$ of regularity at least $\smash{\Cont^1}$. Furthermore, let $h$ be a complete Riemannian metric on $\scrM$ \cite{nomizu-ozeki}, with induced length distance $\smash{\met^h}$, fixed throughout the paper. For $v\in T\scrM$, we write
\begin{align*}
\vert v\vert_h := \sqrt{h(v,v)},
\end{align*}
and we define $\vert v\vert_\Rmet$ analogously provided $\Rmet(v,v)\geq 0$.

We call $v\in T\scrM$ \emph{$\Rmet$-timelike} if $\Rmet(v,v) > 0$, and \emph{$\Rmet$-causal} if $\Rmet(v,v)\geq 0$. Henceforth, we fix a continuous timelike vector field $Z$ on $T\scrM$, and we term $v\in T\scrM\setminus\{0\}$ \emph{future-directed} if $\Rmet(v,Z)>0$, and \emph{past-directed} if $\Rmet(v,Z)<0$. 

A curve  $\gamma\colon [0,1]\to\mms$ is called \emph{future-directed $\Rmet$-timelike}, respectively \emph{future-directed $\Rmet$-causal}, if $\gamma$ is $\smash{\met^h}$-Lipschitz continuous and $\dot{\gamma}_t$ has the respective properties  for $\Leb^1$-a.e.~$t\in[0,1]$. Compared to absolute continuity, Lipschitz continuity is no restriction \cite[p.~17]{minguzzi2019}. We mostly consider the future orientation by $Z$ and hence drop the prefix  ``future-directed'' --- see \autoref{Re:Reversed} below, though --- and, if clear from the context, the metric $\Rmet$ for terminological convenience. 

Let the $\Rmet$-length of a $\Rmet$-causal curve $\gamma\colon[0,1]\to\scrM$ \cite[Def. 5.11]{oneill1983} be given by
\begin{align*}
\Len_\Rmet(\gamma) := \int_0^1\vert \dot{\gamma}_t\vert_\Rmet\d t,
\end{align*}
and define $l_\Rmet\colon\scrM^2\to [0,\infty]\cup\{-\infty\}$ by
\begin{align}\label{Eq:l def}
l_\Rmet(x,y) := \sup\{\Len_\Rmet(\gamma) : \gamma\colon[0,1]\to\scrM\ \Rmet\textnormal{-causal curve},\, \gamma_0= x,\,\gamma_1 = y\},
\end{align}
setting $\sup\emptyset := -\infty$. Slightly deviating from other common definitions --- cf.~e.g. \cite[Def. 2.1]{graf2020} and \autoref{Re:Geod equ} below --- and rather following \cite[Def. 3.27]{kunzinger2018} we use the following notion of geodesics.

\begin{definition} Given $(x,y) \in \smash{l_\Rmet^{-1}([0,\infty])}$, a maximizer $\gamma\colon[0,1]\to\scrM$ of $l_\Rmet(x,y)= \smash{l_\Rmet^+(x,y)}$ in \eqref{Eq:l def} is called \emph{$\Rmet$-geodesic}. 
\end{definition}

For arbitrary sets $C_0,C_1\subset \scrM$, we define
\begin{itemize}
\item the \emph{$\Rmet$-causal future} of $C_0$ by
\begin{align}\label{Eq:HA}
J_\Rmet^+(C_0):= \{y\in \scrM : l_\Rmet(x,y) \geq 0 \textnormal{ for some }x\in C_0\},
\end{align}
\item the \emph{$\Rmet$-causal past} of $C_1$ by
\begin{align}\label{Eq:HAA}
J_\Rmet^-(C_1):= \{x\in \scrM : l_\Rmet(x,y) \geq 0 \textnormal{ for some }y\in C_1\},
\end{align}
\item the \emph{$\Rmet$-causal diamond} of $C_0$ and $C_1$ by
\begin{align*}
    J_\Rmet(C_0,C_1) := J^+(C_0) \cap J^-(C_1).
\end{align*}
\end{itemize}
Given $x,y\in\scrM$ and probability measures $\mu$ and $\nu$ on $\scrM$, we set $\smash{J_\Rmet^\pm(x) := J_\Rmet^\pm(\{x\})}$ and $\smash{J_\Rmet^\pm(\mu) := J_\Rmet^\pm(\supp\mu)}$;  accordingly, we define $\smash{J_\Rmet(x,y)}$ and $\smash{J_\Rmet(\mu,\nu)}$. By replacing ``$\geq$'' by ``$>$'', we define the \emph{$\Rmet$-chronological future} $\smash{I_\Rmet^+(C_0)}$ of $C_0$, the \emph{$\Rmet$-chronological past} $\smash{I_\Rmet^-(C_1)}$ of $C_1$, etc.

We call $(\scrM,\Rmet)$, or simply $\Rmet$, \emph{strongly causal} \cite[Def.~14.11]{oneill1983} if for every $x\in\scrM$ and every open neighborhood $U\subset \scrM$ of $x$, there is  another open neighborhood $V\subset U$ of $x$ such that every $\Rmet$-causal curve with endpoints in $V$ does not leave $U$.  \NEW{}{We term the spacetime $(\scrM,\Rmet)$ \emph{causal} \cite[p.~407]{oneill1983} if it has no closed nonconstant causal curves.}

\begin{definition}\label{Def:GH} The spacetime $(\scrM,\Rmet)$, or simply $\Rmet$, is termed  \emph{globally hyperbolic} if it is causal, and $J_\Rmet(x,y)$ is compact for every $x,y\in\scrM$.
\end{definition}

\NEW{}{In our setting, this definition is equivalent to the traditional definition of global hyperbolicity in terms of  strong causality plus compactness of causal diamonds \cite[Thm.~3.2]{bernal}. In fact, in spacetimes with dimension at least three and $\Rmet$ of class $\smash{\Cont^{1,1}}$, causality can be dropped completely in \autoref{Def:GH} \cite[Thm.~2.7]{hounnonkpe}.}

If not explicitly stated otherwise, in the following we will always assume global hy\-perbolicity of any considered $\Rmet$.

\begin{remark}\label{Re:Geod equ} Important facts used at several occasions inherited by the regularity imposed on $\Rmet$ are the following.
\begin{itemize}
\item Every $\Rmet$-geodesic $\gamma\colon [0,1]\to\scrM$ has a causal character \cite[Prop.~1.2]{lange2020}. More strongly, either $\vert \dot{\gamma}_t\vert_\Rmet > 0$ for every $t\in[0,1]$, or $\vert \dot{\gamma}_t\vert_\Rmet = 0$ for every $t\in[0,1]$. (A similar statement had been obtained before in \cite[Thm. 1.1]{graf2018}, see also \cite[Thm.~2]{sb}.)
\item Every $\Rmet$-geodesic $\gamma\colon [0,1]\to\scrM$  admits a proper-time reparametrization $\smash{\eta\colon[0,1]\to\scrM}$ with regularity $\smash{\Cont^2}$ \cite[Thm.~1.1]{lange2020}, see also \cite[Prop.~2.13]{graf2020} and \cite[Prop.~4.19]{oneill1983}. This means that for every $s,t\in[0,1]$ with $s<t$,
\begin{align}\label{Eq:PTP}
\tsep_\Rmet(\eta_s,\eta_t) = (t-s)\,\tsep_\Rmet(\eta_0,\eta_1).
\end{align}
\end{itemize} 
By the Cauchy--Lipschitz theorem, the latter yields that if the Christoffel symbols of $\Rmet$ are locally Lipschitz continuous, i.e.~provided $\Rmet$ is $\smash{\Cont^{1,1}}$, $\Rmet$-timelike $\Rmet$-geodesics paramet\-rized by proper-time admit no forward or backward branching. That is, if two $\smash{\Cont^2}$-curves $\smash{\eta^1,\eta^2\colon [0,1]\to\scrM}$ arising from the above procedure coincide on some nontrivial subinterval of $[0,1]$, then $\smash{\eta^1=\eta^2}$. In particular, the Lorentzian geodesic space induced by $(\scrM,\Rmet)$ according to \autoref{Sec:Construction} is \emph{$\Rmet$-timelike non\-branching} \cite[Def.~1.10]{cavalletti2020}.
\end{remark}

\subsection{$\Cont^1$-spacetimes as Lorentzian geodesic spaces}\label{Sec:Construction} In this section, following \cite[Sec.~5.1]{kunzinger2018} we review the construction of a \emph{Lorentzian geodesic space}  \cite[Def.~2.8, Def.~3.27]{kunzinger2018} from the given spacetime $\scrM$ with a globally hyperbolic $\smash{\Cont^1}$-Lorentzian metric $\Rmet$. As summarized in \autoref{Pr:Properties} below, this links our setting to the synthetic frameworks of \cite{braun2022,cavalletti2020}. In fact, many of the results in this section hold for merely continuous, strongly causal, and causally plain \cite[Def.~1.16]{chrusciel2012} metrics. Since every Lipschitz metric is causally plain \cite[Cor.~1.17]{chrusciel2012}, we only discuss the case of metric regularity at least $\smash{\Cont^1}$ to streamline the presentation.

 Define two relations $\smash{\ll_\Rmet}$ and $\smash{\leq_\Rmet}$ on $\scrM$ by
 \begin{itemize}
     \item $x\ll_\Rmet y$ if there is a $\Rmet$-timelike curve $\gamma\colon[0,1]\to\scrM$ with $\gamma_0=x$ and  $\gamma_1=y$, or equivalently $\smash{l_\Rmet(x,y)>0}$, and
     \item $x\leq_\Rmet y$ if there is a $\Rmet$-causal curve $\gamma\colon[0,1]\to\scrM$ with $\gamma_0=x$ and  $\gamma_1= y$, or equivalently $\smash{l_\Rmet(x,y)\geq 0}$.
 \end{itemize}
 Given any subset $\mms\subset\scrM$, define 
\begin{align*}
\mms_{\ll_\Rmet}^2 &:= \mms^2 \cap l_\Rmet^{-1}((0,\infty]) = \{(x,y) \in\mms^2 : x\ll_\Rmet y\},\\
\mms_{\leq_\Rmet}^2 &:= \mms^2\cap l_\Rmet^{-1}([0,\infty]) = \{(x,y)\in\mms^2 : x\leq_\Rmet y\}.
\end{align*} 
Clearly, $\ll_\Rmet$ is transitive and contained in $\leq_\Rmet$, i.e.~$\smash{\mms_{\ll_\Rmet}^2 \subset\mms_{\leq_\Rmet}^2}$, and $\smash{\leq_\Rmet}$ is reflexive and transitive, which makes $(\mms,\ll_\Rmet,\leq_\Rmet)$ a \emph{causal space} after \cite[Def.~2.1]{kunzinger2018}.

The positive part $\smash{\tsep_\Rmet := l_\Rmet^+}$ of the function $l_\Rmet$ in \eqref{Eq:l def} is a \emph{time separation function} \cite[Def.~2.8]{kunzinger2018}: it is lower semicontinuous \cite[Prop.~5.7]{kunzinger2018}, and for every $x,y,z\in\scrM$,
\begin{enumerate}[label=\textnormal{\alph*.}]
\item $\smash{\tsep_\Rmet(x,y) = 0}$ provided $x \not\leq_\Rmet y$,
\item $\smash{\tsep_\Rmet(x,y) > 0}$ if and only if $x\ll_\Rmet y$ \cite[Lem.~5.6]{kunzinger2018}, and
\item if $x\leq_\Rmet y \leq_\Rmet z$, we have the \emph{reverse triangle inequality}
\begin{align}\label{Eq:Reverse tau}
\tsep_\Rmet(x,z) \geq \tsep_\Rmet(x,y) + \tsep_\Rmet(y,z).
\end{align}
\end{enumerate}
In particular, the quintuple $(\scrM,\met^h,\ll_\Rmet,\leq_\Rmet,\tsep_\Rmet)$ forms a \emph{Lorentzian pre-length space} \cite[Prop.~5.8]{kunzinger2018} in the sense of \cite[Def.~2.8]{kunzinger2018}.

 Global hyperbolicity of $\Rmet$ entails further fine properties and non-ambiguities of $\smash{(\scrM,\met^h,\ll_\Rmet,\leq_\Rmet,\tsep_\Rmet)}$ as described now. The notion of $\Rmet$-causal curves in \autoref{Sec:Terminology} coincide with the nonsmooth one from \cite[Def.~2.18]{kunzinger2018} (evidently defined solely in terms of $\smash{\leq_\Rmet}$), cf.~\cite[Prop.~5.9]{kunzinger2018}. Moreover, their $\Rmet$-length agrees with their $\smash{\tsep_\Rmet}$-length $\smash{\Len_{\tsep_\Rmet}}$  \cite[Def.~2.24]{kunzinger2018}, cf.~\cite[Rem.~5.1, Lem.~5.10]{kunzinger2018}. In fact, $\smash{(\scrM,\met^h,\ll_\Rmet,\leq_\Rmet, \tsep_\Rmet)}$ is a strongly localizable Lorentz\-ian length space after \cite[Def.~3.16, Def.~3.22]{kunzinger2018}, cf.~\cite[Thm.~5.12]{kunzinger2018}. By the causal ladder for Lorentzian length spaces \cite[Thm.~3.26]{kunzinger2018}, global hyperbolicity of $\Rmet$ after  \autoref{Def:GH} is then equivalent to global hyperbolicity of $\smash{(\scrM,\met^h,\ll_\Rmet,\leq_\Rmet,\tsep_\Rmet)}$  \cite[Def.~2.35]{kunzinger2018}, i.e.~we have 
\begin{enumerate}[label=\textnormal{\alph*.}]
\item compactness of causal diamonds between any $x,y\in\scrM$, and 
\item \emph{non-total imprisonment}, i.e.~for every compact $C\subset\scrM$,
\begin{align}\label{Eq:NTI}
\sup\{\Len_{\met^h}(\gamma) : \gamma\colon[0,1]\to\scrM\ \Rmet\text{-causal curve},\, \gamma_{[0,1]}\subset C\} < \infty.
\end{align}
\end{enumerate}
In particular, $\smash{\tsep_\Rmet}$ is finite and continuous \cite[Thm.~3.28]{kunzinger2018}.

Lastly, a combination of \cite[Prop.~3.3, Cor.~3.4]{saemann2016} with \cite[Thm.~3.26, Thm.~3.28]{kunzinger2018} and \autoref{Re:Geod equ} gives that $\smash{(\scrM,\met^h,\ll_\Rmet,\leq_\Rmet,\tsep_\Rmet)}$ satisfies all regularity properties required for the most important synthetic results in \cite{braun2022, cavalletti2020} as follows.

\begin{proposition}\label{Pr:Properties} The space $\smash{(\scrM,\met^h,\ll_\Rmet,\leq_\Rmet,\tsep_\Rmet)}$ is a regular Lorentzian length space \cite[Def.~3.16, Def.~3.22]{kunzinger2018} with the following properties.
\begin{enumerate}[label=\textnormal{(\roman*)}]
\item \textnormal{\textbf{Causal closedness.}} The set $\smash{\scrM_{\leq_\Rmet}^2}$ is closed in $\scrM^2$.
\item \textnormal{\textbf{$\boldsymbol{\scrK}$-global hyperbolicity.}} For every compact $C_0,C_1 \subset \scrM$, the causal dia\-mond $\smash{J_\Rmet(C_0,C_1)}$ is compact in $\scrM$.
\item \textnormal{\textbf{Geodesy.}} Every $x,y\in\scrM$ with $\smash{x\leq_\Rmet y}$ are joined by a $\Rmet$-geodesic.
\end{enumerate}
\end{proposition}

\subsection{Approximation}\label{Sub:Approx}  Since $g$ is of class $\smash{\Cont^1}$ and since curvature quantities involve second derivatives of the metric components, these have to be  defined distributionally in a sense that we briefly recall from \cite[Sec.~3]{graf2020}, see also \cite{grosser2001,kunzinger2022,steinbauer2008}. 

\subsubsection{Distributional curvature bounds}\label{Sub:DIST} The space of \emph{distributions} $\scrD'(\scrM)$ is defined as the topological dual of the space of smooth, compactly supported sections of the volume bundle $\mathrm{Vol}(\scrM)$, i.e.
\begin{align*}
\scrD'(\scrM):=\Gamma_\rmc(\mathrm{Vol}(\scrM))'.
\end{align*}
An element $\mu\in\smash{\Gamma_\rmc(\mathrm{Vol}(\scrM))}$ is called \emph{volume density}. The pairing of $u\in\scrD'(\scrM)$ with $\mu$ will be denoted $\langle u,\mu\rangle$. 

We naturally regard $\smash{\Cont^\infty(\scrM)}$ as subspace of $\scrD'(\scrM)$ by identification of a given $f\in\Cont^\infty(\scrM)$ with the functional $\mu \mapsto \int_{\scrM} f\,\mu$ on $\smash{\Gamma_\comp(\textnormal{Vol}(\scrM))}$.

The above definition can be generalized to \emph{tensor distributions}. More precisely, given $r,s\in\N_0$ the space of $\smash{T_s^r\scrM}$-valued distributions --- with $r$ covariant and $s$ contravariant slots --- is defined by
\begin{align*}
\scrD'\scrT_s^r(\scrM) := \Gamma_\rmc(T_r^s\scrM\otimes \textnormal{Vol}(\scrM))' \cong \scrD'(\scrM) \otimes_{\Cont^\infty(\scrM)} T_s^r\scrM.
\end{align*}
In particular, every tensor distribution is locally defined by its proper coefficients in $\scrD'(\scrM)$. That is, for a given atlas $(U_\alpha,\psi_\alpha)_{\alpha\in A}$, the restriction $\smash{T\big\vert_{U_\alpha}}$ of $T\in\scrD'\scrT_s^r(\scrM)$ to $U_\alpha$ can be written as
\begin{align}\label{Eq:Coeff repr}
T\big\vert_{U_\alpha} = (^\alpha T)_{j_1\dots j_s}^{i_1\dots i_r}\,\frac{\partial}{\partial x^{i_1}} \otimes \dots \otimes \frac{\partial}{\partial x^{i_r}} \otimes \rmd x^{j_1} \otimes \dots \otimes \rmd x^{j_s}
\end{align}
using Einstein's summation convention, with local coefficients $\smash{(^\alpha T)_{j_1\dots j_s}^{i_1\dots i_r} \in \scrD'(U_\alpha)}$. Via the chart map $\psi_\alpha$, the latter can both be pushed forward to and recovered by pullback from a distribution on $\R^n$; cf.~\cite[Prop.~3.1]{graf2020} for details. 

In view of the next definition \cite[Def.~3.2]{graf2020}, we call $\mu\in \Gamma_\comp(\textnormal{Vol}(\scrM))$  \emph{nonnegative} provided $\smash{\int_U\mu \geq 0}$ for every open $U\subset\scrM$.

\begin{definition}\label{Def:Nonneg} Let $u\in\scrD'(\scrM)$. We write $u\geq0$  if $\langle u,\mu\rangle\geq 0$ for every nonnegative volume density $\mu \in \Gamma_\comp(\textnormal{Vol}(\scrM))$. Analogously, given any $v\in\scrD'(\scrM)$ we write $u\geq v$ provided $u-v\geq 0$.
\end{definition}

Given the $\smash{\Cont^1}$-metric $\Rmet$ with Christoffel symbols $\smash{\Gamma_{ij}^k}$, a smooth vector field $X$ over $\scrM$ with local components $\smash{X^1,\dots,X^n}$, some $V\in\Cont^1(\scrM)$, and $N\in[n,\infty)$, the following quantities are locally well-defined in $\scrD'(\scrM)$ by the usual formulas:
\begin{align*}
\Ric_\Rmet(X,X) &:= \Big[\frac{\partial\Gamma_{ij}^m}{\partial x^m} - \frac{\partial \Gamma_{im}^m}{\partial x^j} + \Gamma_{ij}^m\,\Gamma_{km}^k - \Gamma_{ik}^m\,\Gamma_{jm}^k\Big]\,X^i\,X^j,\\
\Hess_\Rmet V(X,X) &:= \Big[\frac{\partial^2 V}{\partial x^i\partial x^j} - \Gamma_{ij}^k\,\frac{\partial V}{\partial x^k}\Big]\,X^i\,X^j,\\
\Ric_\Rmet^{N,V}(X,X) &:= \Ric_\Rmet(X,X) + \Hess_\Rmet V(X,X) - \frac{1}{N-n}\,\rmD V(X)^2.
\end{align*}
If $N = n$, we assume $V$ to be constant by default, so that $\rmD V(X)^2/(N-n) := 0$. Evidently, these definitions give rise to nonrelabeled tensor distributions 
\begin{align*}
\Ric_\Rmet,\, \Hess_\Rmet V,\, \Ric_\Rmet^{N,V}\in \scrD'\scrT_2^0(\scrM).
\end{align*}

\begin{definition}\label{Def:Curv bounds} Given $V$ and $N$ as above and any $K\in\R$, we say
\begin{align*}
\Ric_\Rmet^{N,V} \geq K\text{ in all timelike directions}
\end{align*}
if for every smooth $\Rmet$-timelike vector field $X$ on $\scrM$,
\begin{align}\label{Eq:Def inequ}
\Ric_\Rmet^{N,V}(X,X) \geq K\,\big\vert X\big\vert_\Rmet^2
\end{align}
holds in the sense of \autoref{Def:Nonneg}.
\end{definition}

\begin{remark}\label{RE:T} If $\Rmet$ and $V$ are of class $\smash{\Cont^{1,1}}$, $\smash{\Ric_\Rmet^{N,V}(X,X)}$ is well-defined as an element of $\smash{\Ell_\loc^\infty(\scrM,\vol_\Rmet)}$, cf.~\autoref{Sub:MT}. In this case, the condition 
\begin{align*}
\Ric_\Rmet^{N,V} \geq K\textnormal{ in all timelike directions}
\end{align*}
holds if and only if for every $X$ as in \autoref{Def:Curv bounds}, \eqref{Eq:Def inequ} holds $\vol_\Rmet$-a.e.; if $\Rmet$ and $V$ are of class $\smash{\Cont^2}$, this characterization improves to a pointwise statement of \eqref{Eq:Def inequ}. 
\end{remark}

\subsubsection{Regularization of the metric} Now we show how to approximate $\Rmet$ in a ``nice'' way. That is, if $\Rmet$ obeys   distributional curvature bounds, recall \autoref{Def:Curv bounds}, it will even be possible to almost preserves these bounds, at least locally, cf.~\autoref{Le:Approx}.

In order to approximate $\Rmet$, we need to clarify how to regularize a distribution over $\scrM$. Fix a standard mollifier $\{\rho_\varepsilon : \varepsilon > 0\}$ in $\R^n$, a countable atlas $(U_\alpha,\psi_\alpha)_{\alpha\in\N}$ with relatively compact $U_\alpha$, a subordinate partition of unity $(\xi_\alpha)_{\alpha\in \N}$, as well as functions $\chi_\alpha\in\Cont_\comp^\infty(U_\alpha)$ with $\vert\chi_\alpha\vert(\scrM) = [0,1]$ and $\chi_\alpha = 1$ on an open neighborhood of $\supp\xi_\alpha$ in $U_\alpha$. 

As usual, the convolution of a Euclidean distribution $u$ with compact support \cite[p.~1434]{graf2020} in an open set $\Omega\subset\R^n$ with $\rho_\varepsilon$, $\varepsilon \in (0, \met_\rmE(\supp u, \partial \Omega))$, is the smooth function $u\star\rho_\varepsilon$ on $\Omega$ given by
\begin{align}\label{Eq:Distr}
(u\star\rho_\varepsilon)(x) := \langle u,\rho_\varepsilon(x-\cdot)\rangle.
\end{align}
Then for $T\in\scrD'\scrT_s^r(\scrM)$ we define a smooth $(r, s)$-tensor field $T\star_\scrM\rho_\varepsilon$ by
\begin{align*}
T\star_\scrM\rho_\varepsilon:= \sum_{\alpha\in\N}\chi_\alpha\,(\psi_\alpha^{-1})_*\big[(\psi_\alpha)_*(\xi_\alpha T)\star\rho_\varepsilon\big],
\end{align*}
where the convolution on the right-hand side is understood componentwise in terms of the push-forwards of the local coefficients from \eqref{Eq:Coeff repr} to $\R^n$ via \eqref{Eq:Distr}.

Clearly, for every $u\in\scrD'(\scrM)$ and every $\varepsilon > 0$, we have $u\star_\scrM \rho_\varepsilon \geq 0$ if $u\geq 0$.

In the most relevant case for our purposes, namely  $T := g\in \scrD'\scrT_2^0(\scrM)$, it follows from basic properties of mollification in $\R^n$ \cite[Prop.~3.5]{graf2020} that $\Rmet \star_\scrM\rho_\varepsilon \to \Rmet$ in $\smash{\Cont_\loc^1(\scrM)}$ as $\varepsilon \to 0$. However, this convergence is too weak to ensure mollification of $\Rmet$ to (almost) preserve distributional curvature bounds. Neither are light cones with respect to $\Rmet\star_\scrM\rho_\varepsilon$ \NEW{``thinner''}{``narrower''} than those of $\Rmet$, a property which will be used multiple times below. Both issues are resolved in the following crucial \autoref{Le:Approx} which summarizes \cite[Lem.~4.1, Lem.~4.2, Lem.~4.5]{graf2020}. (The arguments therein can easily be adapted to cover the case of arbitrary curvature lower bounds $K\in\R$ and arbitrary $\smash{\Cont^1}$-weights $V$, cf.~\cite[Thm.~4.3, Rem.~4.4]{kunzinger2022}.)

Recall that for a Lorentzian metric $\tilde{\Rmet}$,  by $\tilde{\Rmet}\prec \Rmet$ we mean that every $\tilde{\Rmet}$-causal tangent vector $v$ is $\Rmet$-timelike (more visually, that $\tilde{\Rmet}$-light cones are strictly \NEW{``thinner''}{``narrower''} than $\Rmet$-light cones). Also, for $\smash{T\in \scrD'\scrT_2^0(\scrM)}$ and a compact $C\subset \scrM$ we set
\begin{align*}
\Vert T\Vert_{\infty,C} := \sup\{\vert T(x)(v,w)\vert : x\in C,\,v,w\in T_x\scrM \textnormal{ with }\vert v\vert_h = \vert w\vert_h = 1\}.
\end{align*}
Lastly, let $^h\nabla$ denote the Levi-Civita connection with respect to $h$.

\begin{lemma}\label{Le:Approx} There exist \emph{smooth} Lorentzian metrics $\{\check{\Rmet}_\varepsilon : \varepsilon > 0\}$ on $\scrM$, time-ori\-en\-table by the same timelike vector field $Z$ as $\Rmet$, with the following properties.
\begin{enumerate}[label=\textnormal{\textcolor{black}{(}\roman*\textcolor{black}{)}}]
    \item\label{zrs} We have $\check{g}_\varepsilon\prec g$ for every $\varepsilon>0$.
    \item We have $\smash{\check{g}_\varepsilon-g\star_\scrM\rho_\varepsilon\to 0}$ in $\Cont^\infty_\loc(\scrM)$ as $\varepsilon \to 0$. That is, for every compact $C\subset\scrM$ and every $i\in \N$, we have
    \begin{align*}
    \lim_{\varepsilon \to 0} \big\Vert ({^h}\nabla)^i \check{\Rmet}_\varepsilon - ({^h}\nabla)^i(\Rmet\star_\scrM\rho_\varepsilon) \big\Vert_{\infty,C} = 0.
    \end{align*}
    In particular, $\check{g}_\varepsilon \to g$ in $\Cont^1_\loc(\scrM)$ as $\varepsilon \to 0$, i.e.~for every compact $C\subset\scrM$ and every $i\in\{0,1\}$,
    \begin{align*}
    \lim_{\varepsilon\to 0} \big\Vert (^h\nabla)^i\check{\Rmet}_\varepsilon - (^h\nabla)^i\Rmet \big\Vert_{\infty,C}=0.
    \end{align*}
\end{enumerate}

Moreover, let $\smash{V\in \Cont^1(\scrM)}$ and $N\in [n,\infty)$, and assume 
\begin{align*}
\Ric_{\Rmet}^{N,V} \geq K \text{ in all timelike directions}.
\end{align*} 
Then $\{\check{\Rmet}_\varepsilon : \varepsilon  >0\}$ can be constructed to have the following further property. For every compact $C\subset \scrM$ and every $c,\delta,\kappa>0$, there exists $\varepsilon_0 > 0$ such that for every $\varepsilon \in (0,\varepsilon_0)$ and every $\smash{v\in T\scrM\big\vert_C}$, we have
\begin{align*}
\vert v\vert_{\check{\Rmet}_\varepsilon} \geq \sqrt{\kappa},\ \vert v\vert_h \leq \sqrt{c} \quad\Longrightarrow\quad 
\Ric_{\check{\Rmet}_\varepsilon}^{N,V}(v,v) \geq (K-\delta)\,\big\vert v\big\vert_{\check{\Rmet}_\varepsilon}^2.
\end{align*}
\end{lemma}

Knowledge of the following consequence of \ref{zrs} above and \cite[Rem.~1.1]{graf2020} will be relevant e.g.~in the proofs of \autoref{Le:tauinfty tauk}, and \autoref{Pr:Displacement}.

\begin{lemma}\label{Le:Glob hyperb} For every $\varepsilon>0$, $\check{\Rmet}_\varepsilon$ is globally hyperbolic.
\end{lemma}

\subsection{Lorentzian optimal transport}\label{Sub:Lor opt transport} Lastly, we recall basic elements of Lorentz\-ian optimal transport theory, referring to \cite{cavalletti2020, eckstein2017, kell2020, mccann2020, mondinosuhr2018, suhr2018} for details. 

Evidently, all subsequent notions with background space $\scrM$ will make sense on any closed subset $\mms\subset\scrM$.

\subsubsection{Measure-theoretic notation}\label{Sub:MT} Let $\scrP(\scrM)$ be the class of  Borel probability measures on $\scrM$, and let $\scrP_\comp(\scrM)$ consist of all $\mu\in\scrP(\scrM)$ with compact \emph{support} $\supp\mu\subset\scrM$. 

Let $\vol_\Rmet$ be the Lo\-rentz\-ian volume measure on $\scrM$ associated to $\Rmet$. It arises from the  volume form $\rmd\vol_\Rmet$ induced by $\Rmet$ by the formula
\begin{align}\label{Eq:Local repr}
\rmd \vol_\Rmet\big\vert_U := \sqrt{\vert\!\det \Rmet\vert}\d x^1\wedge\dots\wedge\rmd x^n
\end{align}
on a coordinate chart $(U,\psi)$, where $\smash{\{\rmd x^1(x),\dots,\rmd x^n(x)\}}$ is a positively oriented basis of $T_x^*\scrM$ for every $x\in U$. Let $\scrP^\ac(\scrM,\vol_\Rmet)$ be the set of all $\vol_\Rmet$-absolutely continuous elements of $\scrP(\scrM)$, and set $\scrP_\comp^\ac(\scrM,\vol_\Rmet) := \scrP_\comp(\scrM)\cap\scrP^\ac(\scrM,\vol_\Rmet)$.

Given $\mu,\nu\in\scrP(\scrM)$, let $\Pi(\mu,\nu)$ be the set of all their  \emph{couplings}, i.e.~all $\pi\in\scrP(\scrM^2)$ such that $\pi[\,\cdot\times\scrM] = \mu$ and $\pi[\scrM\times\cdot\,]=\nu$. This concept of couplings conveniently makes sense of chronology and causality relations between $\mu$ and $\nu$ in terms of their supports, namely in terms of the sets $\smash{\Pi_{\ll_\Rmet}(\mu,\nu)}$ and $\smash{\Pi_{\leq_\Rmet}(\mu,\nu)}$, respectively, which consist of all $\pi\in\Pi(\mu,\nu)$ with $\smash{\pi[\scrM_{\ll_\Rmet}^2]=1}$ and $\smash{\pi[\scrM_{\leq_\Rmet}^2]=1}$, respectively.

\subsubsection{The $\smash{\ell_{\Rmet,p}}$-optimal transport problem} Given $p\in(0,1)$, define $\smash{\ell_{\Rmet,p}}\colon \scrP(\scrM)^2 \to [0,\infty]\cup\{-\infty\}$ through
\begin{align}\label{Eq:lp def}
\ell_{\Rmet,p}(\mu,\nu) := \sup\{\Vert \tsep_\Rmet\Vert_{\Ell^p(\scrM^2,\pi)} : \pi \in\Pi_{\leq_\Rmet}(\mu,\nu)\},
\end{align}
subject to the usual convention $\smash{\ell_{\Rmet,p}(\mu,\nu) := -\infty}$ if $\smash{\Pi_{\leq_\Rmet}(\mu,\nu) = \emptyset}$. 
This quantity is morally interpreted as a time separation function on $\scrP(\scrM)$, compare with \eqref{Eq:Reverse tau}: indeed  \cite[Prop.~2.5]{cavalletti2020}, for every $\mu,\nu,\sigma\in\scrP(\scrM)$,
\begin{align*}
\ell_{\Rmet,p}(\mu,\sigma) \geq \ell_{\Rmet,p}(\mu,\nu) + \ell_{\Rmet,p}(\nu,\sigma).
\end{align*}
Given $\mu,\nu\in\scrP(\scrM)$, we call $\pi\in\Pi(\mu,\nu)$ \emph{$\smash{\ell_{\Rmet,p}}$-optimal} if $\smash{\pi\in\Pi_{\leq_\Rmet}(\mu,\nu)}$ and $\pi$ realizes the supremum in  \eqref{Eq:lp def}. Concerning existence of such $\pi$, for our purposes it will suffice to know that if $\mu,\nu\in\scrP_\comp(\scrM)$ with $\smash{\Pi_{\leq_\Rmet}(\mu,\nu)\neq\emptyset}$ --- a condition which holds for $\pi := \mu\otimes\nu$ if $\smash{\supp\mu\times\supp\nu\subset\scrM_{\leq_\Rmet}^2}$ --- admit an $\smash{\ell_{\Rmet,p}}$-optimal coupling; also, by compactness of $\supp\mu\times\supp\nu$ we clearly have
\begin{align*}
\ell_{\Rmet,p}(\mu,\nu) \leq \sup\tsep_\Rmet(\supp\mu\times\supp\nu)  < \infty.
\end{align*}

In view of our intended synthetic treatment of  \emph{$\Rmet$-timelike} Ricci curvature bounds we recall the following definition by \cite[Def.~2.18]{cavalletti2020}, see also \cite[Def.~4.1]{mccann2020}.

\begin{definition}\label{Def:TL p dual} We call a pair $\smash{(\mu,\nu)\in\scrP_\comp(\scrM)^2}$ \emph{$\Rmet$-timelike $p$-dualizable} if
\begin{align*}
\{\pi\in\Pi(\mu,\nu) : \pi\text{ is } \ell_{\Rmet,p}\textnormal{-optimal}\}\cap \Pi_{\ll_\Rmet}(\mu,\nu)\neq\emptyset.
\end{align*}
Any element of the set on the left-hand side is called \emph{$\Rmet$-timelike $p$-dualizing}.
\end{definition}

\begin{remark}\label{Re:DUL}
By the preceding discussion, it is evident that if $\smash{\mu,\nu\in\scrP_\comp(\scrM)}$ satisfy $\smash{\supp\mu\times\supp\nu\subset\scrM_{\ll_\Rmet}^2}$, then the pair $(\mu,\nu)$ is $\Rmet$-timelike $p$-dualizable (even in a stronger sense \cite[Def.~2.27]{cavalletti2020}, cf.~\cite[Cor.~2.29]{cavalletti2020}).
\end{remark}

\subsubsection{Timelike proper-time parametrized $\smash{\ell_{\Rmet,p}}$-geodesics} Next, we review the technical definition of geodesics with respect to $\smash{\ell_{\Rmet,p}}$, referring to \cite[Subsec.~2.3.6, App.~B]{braun2022} for details. The idea is to construct the latter as ``proper-time reparametrizations'' of plans concentrated on $\Rmet$-geodesics, i.e.~$\smash{\Len_\Rmet}$-maximizing $\Rmet$-causal curves. Compared to the weaker notion of timelike $\smash{\ell_{\Rmet,p}}$-geodesics from \cite[Def.~1.1]{mccann2020}, in a more general synthetic setting this procedure allows for good compactness properties  more evidently \cite[Prop.~B.11]{braun2022}, as implicitly used many times in \autoref{Ch:Proofs}. \NEW{}{(Although with some more effort, it should be possible to prove the notion from \cite{mccann2020} has similar properties.)} If $\Rmet$ is smooth, no ambiguity occurs in all relevant cases \cite[Rem.~B.10]{braun2022}.

Let $\smash{\Geo_\Rmet(\scrM)}$ be the set of $\Rmet$-geodesics $\gamma\colon[0,1]\to\scrM$; it is Polish by \autoref{Pr:Properties} and non-total imprisonment, cf.~\eqref{Eq:NTI}. Furthermore, let $\eval_t\colon\Geo_\Rmet(\scrM) \to \scrM$ be the evaluation map $\eval_t(\gamma) := \gamma_t$, $t\in[0,1]$. Set
\begin{align*}
\TGeo_\Rmet(\scrM) := \{\gamma\in\Geo_\Rmet(\scrM) : \tsep_\Rmet(\gamma_0,\gamma_1)>0\},
\end{align*}
which precisely consists of \emph{$\Rmet$-timelike} $\Rmet$-geodesics by \autoref{Re:Geod equ}. By the proof of \cite[Prop.~9.1]{lange2020}, see also \cite[Lem.~B.6]{braun2022} and \cite[Cor.~3.35]{kunzinger2018}, there exists a continuous reparametrization map $\sfr\colon \TGeo_\Rmet(\scrM)\to \Cont([0,1];\scrM)$ such that $\eta := \sfr(\gamma)$ obeys \eqref{Eq:PTP} for every $\gamma\in\TGeo_\Rmet(\scrM)$. With this said, given $\mu_0,\mu_1\in\scrP(\scrM)$ we set
\begin{align*}
\OptTGeo_{\ell_{\Rmet,p}}^{\tsep_\Rmet}(\mu_0,\mu_1) &:= \sfr_\push\{\bdpi\in\scrP(\Geo_\Rmet(\scrM)) : (\eval_0,\eval_1)_\push\bdpi  \text{ is }\ell_{\Rmet,p}\text{-optimal}\\ 
&\qquad\qquad \textnormal{with }(\eval_0,\eval_1)_\push\bdpi[\scrM_{\ll_\Rmet}^2]=1\}.
\end{align*}

\begin{definition} A curve $(\mu_t)_{t\in[0,1]}$ in $\scrP(\scrM)$ is called \emph{timelike proper-time paramet\-rized $\smash{\ell_{\Rmet,p}}$-geodesic} if it is represented by some $\smash{\bdpi\in\OptTGeo_{\ell_{\Rmet,p}}^{\tsep_\Rmet}(\mu_0,\mu_1)}$, i.e.
\begin{align*}
\mu_t= (\eval_t)_\push\bdpi
\end{align*}
for every $t\in[0,1]$; such  a $\bdpi$ is called \emph{timelike $\smash{\ell_{\Rmet,p}}$-optimal geodesic plan}.
\end{definition}

By construction, every timelike $\smash{\ell_{\Rmet,p}}$-optimal geodesic plan $\bdpi$ is concentrated on $\Rmet$-causal curves which satisfy \eqref{Eq:PTP}. As a corollary of \eqref{Eq:Reverse tau}, every timelike proper-time parametrized $\smash{\ell_{\Rmet,p}}$-geodesic $(\mu_t)_{t\in[0,1]}$ is a timelike $\smash{\ell_{\Rmet,p}}$-geodesic in the sense of \cite[Def.~1.1]{mccann2020} if $\smash{\ell_{\Rmet,p}(\mu_0,\mu_1)<\infty}$: indeed, for every $s,t\in[0,1]$ with $s<t$,
\begin{align*}
\ell_{\Rmet,p}(\mu_s,\mu_t) = (t-s)\,\ell_{\Rmet,p}(\mu_0,\mu_1)\in (0,\infty).
\end{align*}

\subsubsection{Synthetic timelike lower Ricci curvature bounds} The subsequent synthetic definitions of timelike Ricci curvature lower bounds --- foreshadowed by the works \cite{cavalletti2020, mccann2020, mondinosuhr2018} which studied a different entropy functional --- have been set up for general measured Lorentzian spaces \cite[Def.~1.11]{cavalletti2020} in \cite[Def.~3.3, Def. 4.1]{braun2022}. These constitute  Lorentzian counterparts of analogous notions for metric measure spaces, cf.~\cite[Def.~2.1]{ohta2007} and \cite[Def.~1.3, Def.~5.1]{sturm2006b}. 

This is where a reference measure comes into play: given $V\in\Cont^1(\scrM)$, set
\begin{align*}
\mmeas_\Rmet^V := \rme^{-V}\,\vol_\Rmet.
\end{align*} 
The associated measured Lorentzian structure, recall \autoref{Sec:Construction}, is written
\begin{align}\label{Eq:X def}
\scrX_\Rmet^V := (\scrM,\met^h,\mmeas_\Rmet^V, \ll_\Rmet,\leq_\Rmet,\tsep_\Rmet).
\end{align}
For $N\in[1,\infty)$, subject to the Lebesgue decomposition $\smash{\mu = \rho\,\mmeas_\Rmet^V + \mu_\perp}$ of $\mu\in\scrP(\scrM)$, the \emph{$N$-Rényi entropy} $\scrS_{\Rmet}^{N,V}\colon \scrP(\scrM)\to [-\infty,0]$ with respect to $\smash{\mmeas_\Rmet^V}$ is 
\begin{align}\label{Eq:Renyi}
\scrS_{\Rmet}^{N,V}(\mu) := -\int_\scrM \rho^{-1/N}\d\mu = -\int_\scrM\rho^{1-1/N}\d\mmeas_\Rmet^V.
\end{align}
If $V=0$ and $N=n$, it reduces to the $n$-Rényi entropy $\smash{\scrS_\Rmet^n}$ defined in \autoref{Ch:Intro}.

Moreover, for $t\in[0,1]$ and $K\in\R$,  we define the \emph{distortion coefficients} $\smash{\tau_{K,N}^{(t)}}$ \cite[p.~137]{sturm2006b} as follows. Given any $\vartheta\geq 0$, set
\begin{align}\label{Eq:Distortion coeff}
\begin{split}
\mathfrak{s}_{K,N}(\vartheta) &:= \begin{cases} \displaystyle\frac{\sin(\sqrt{KN^{-1}}\,\vartheta)}{\sqrt{KN^{-1}}}\vspace*{0.05cm} & \text{if }K>0,\\
\vartheta & \text{if }K=0,\\
\displaystyle\frac{\sinh(\sqrt{-KN^{-1}}\,\vartheta)}{\sqrt{-KN^{-1}}} & \text{otherwise},
\end{cases}\\
\sigma_{K,N}^{(t)}(\vartheta) &:= \begin{cases} \infty\vspace*{0.15cm} & \text{if }K\vartheta^2 \geq N\pi^2,\\
t & \textnormal{if }K\vartheta^2=0,\\
\displaystyle \frac{\mathfrak{s}_{K,N}(t\,\vartheta)}{\mathfrak{s}_{K,N}(\vartheta)} & \text{otherwise},
\end{cases}\\
\tau_{K,N}^{(t)}(\vartheta) &:= t^{1/N}\,\sigma_{K,N-1}^{(t)}(\vartheta)^{1-1/N}.
\end{split}
\end{align}
We always have the inequality
\begin{align*}
\sigma_{K,N}^{(t)}(\vartheta)\leq \tau_{K,N}^{(t)}(\vartheta).
\end{align*}
In the nondegeneracy cases $K\,\vartheta^2 =0$ or $K\,\vartheta^2 < N\,\pi^2$, the function $u\colon[0,1]\to\R$ given by $\smash{u(t) := \sigma_{K,N}^{(t)}(\vartheta)}$ is the unique solution to the ODE
\begin{align*}
u''(t) + \frac{K}{N}\,\vartheta^2\,u(t) = 0
\end{align*}
with boundary data $u(0)=0$ and $u(1)=1$. The quantity $\smash{\tau_{K,N}^{(t)}(\vartheta)}$ is a geometric average of two distortion coefficients. Roughly speaking, it encodes the behavior of the optimal transport Jacobian in timelike directions: it is affected by curvature in ``$N-1$'' directions orthogonal to the transport, represented by $\smash{\sigma_{K,N-1}^{(t)}(\vartheta)}$, while the contribution tangential to the transport does not see any curvature, represented by $t$. Compare with the proof of \cite[Thm.~5.9]{braunohta}.

\begin{definition}\label{Def:TCD} Let $p\in (0,1)$, $K\in\R$, and $N\in[1,\infty)$. We say $\smash{\scrX_\Rmet^V}$ satisfies  the \emph{timelike curvature-dimension condition} $\TCD_p(K,N)$ if for every $\Rmet$-timelike $p$-dualizable pair $\smash{(\mu_0,\mu_1) = (\rho_0\,\mmeas_\Rmet^V, \rho_1\,\mmeas_\Rmet^V)\in\scrP_\comp^\ac(\scrM,\vol_\Rmet)}$, there exist 
\begin{itemize}
\item a timelike proper-time parametrized $\ell_{\Rmet,p}$-geodesic $(\mu_t)_{t\in[0,1]}$ connecting $\mu_0$ to $\mu_1$, and 
\item a $\Rmet$-timelike $p$-dualizing coupling $\pi\in\smash{\Pi_{\ll_\Rmet}(\mu_0,\mu_1)}$ 
\end{itemize}
such that for every $t\in[0,1]$ and every $N'\geq N$,
\begin{align*}
\scrS_{\Rmet}^{N',V}(\mu_t) &\leq -\int_{\scrM^2} \tau_{K,N'}^{(1-t)}(\tsep_\Rmet(x^0,x^1))\,\rho_0(x^0)^{-1/N'}\d\pi(x^0,x^1)\\
&\qquad\qquad - \int_{\scrM^2} \tau_{K,N'}^{(t)}(\tsep_\Rmet(x^0,x^1))\,\rho_1(x^1)^{-1/N'}\d\pi(x^0,x^1).
\end{align*}
\end{definition}

\begin{definition}\label{Def:TMCP} Let\NEW{$p\in (0,1)$,}{} $K\in\R$, and $N\in[1,\infty)$. We say $\smash{\scrX_\Rmet^V}$ satisfies the \emph{timelike measure-contraction property} $\TMCP(K,N)$ if for every $\smash{\mu_0 =\rho_0\,\mmeas_\Rmet^V}\in\smash{\scrP_\comp^\ac(\scrM,\vol_\Rmet)}$ and every $x_1 \in \scrM$ with $\smash{\mu_0[I_\Rmet^-(x_1)]=1}$, there exist $p\in (0,1)$ and a timelike proper-time parame\-tri\-zed $\ell_{\Rmet,p}$-geodesic $(\mu_t)_{t\in[0,1]}$ from $\mu_0$ to $\smash{\mu_1 := \delta_{x_1}}$  such that for every $t\in [0,1)$ and every $N'\geq N$,
\begin{align*}
\scrS_{\Rmet}^{N',V}(\mu_t) \leq -\int_{\scrM} \tau_{K,N'}^{(1-t)}(\tsep_\Rmet(x^0,x_1))\,\rho_0(x^0)^{-1/N'}\d\mu_0(x^0).
\end{align*}
\end{definition}

These conditions are compatible with the smooth case, in the sense that if $\Rmet$ is smooth, roughly speaking, $\smash{\TCD_p(K,N)}$ and $\smash{\TMCP(K,N)}$  characterize  $\Rmet$-timelike Ricci curvature lower bounds by $K\in\R$ and upper dimension bounds by $N\in[1,\infty)$ for $\smash{(\scrM,\Rmet)}$ \NEW{\cite[Thm.~A.1, Thm.~A.5]{braun2022}}{\cite[Thm.~5.9, Thm.~6.1]{braunohta}}; \autoref{Th:TMCP} and \autoref{Th:TCD nonbr} will extend these results to lower regularity Lorentzian metrics. (The results from \cite{braunohta} are even proven for more general smooth Finsler spacetimes.)

\NEW{}{We note that $\TMCP(K,N)$ does not depend on the given transport exponent $p$ in \autoref{Def:TMCP}. Indeed, the collection  $(\mu_t)_{t\in[0,1]}$ therein is a timelike proper-time parametrized $\smash{\ell_{\Rmet,p}}$-geodesic if and only if it constitutes a timelike proper-time pa\-ra\-metrized $\smash{\ell_{\Rmet,p'}}$-geodesic for \emph{every} $p'\in (0,1)$. This simply follows because $\smash{\Pi_{\ll_\Rmet}(\mu_0,\mu_1)} = \Pi(\mu_0,\mu_1)$ is a singleton; cf.~\cite[Rem.~4.3]{braun2022} and \cite[Rem.~2.4]{cav}.}

Moreover, the following basic properties hold. 

\begin{itemize}
\item Both notions are consistent in the ``curvature parameter'' $K$ and the ``dimensional parameter'' $N$ \cite[Prop.~3.7, Prop.~4.5]{braun2022}. 
\item Moreover, $\smash{\TCD_p(K,N)}$ implies $\smash{\TMCP(K,N)}$ \cite[Prop.~4.8]{braun2022}, yet the latter condition is strictly weaker in general  \cite[Rem.~A.5]{braun2022}. 
\end{itemize}

\begin{remark}\label{Re:Relations} The condition $\smash{\TCD_p(K,N)}$ implies the reduced timelike curvature-dimension condition $\smash{\TCD_p^*(K,N)}$ from \cite[Def.~3.2]{braun2022}, cf.~\cite[Prop.~3.6]{braun2022}. Under $\Rmet$-timelike nonbranching according to \autoref{Re:Geod equ}, the latter condition is   \emph{equivalent} to the $\smash{\TCD_p^e(K,N)}$ condition introduced in \cite[Def.~3.2]{cavalletti2020} after \cite{mccann2020, mondinosuhr2018}, which is formulated in terms of the \emph{Boltzmann} entropy, by \cite[Thm.~3.35]{braun2022}. Analogous chains  of implications are satisfied by $\smash{\TMCP(K,N)}$ \cite[Prop.~4.5, Thm. 4.20]{braun2022}.

Motivated by an analogous result for (essentially) nonbranching metric measure spaces \cite{cavmon}, all three timelike curvature-dimension conditions are in fact expected to coincide under appropriate timelike nonbranching hypotheses.
\end{remark}

\begin{remark}\label{Re:Reversed} Starting from $(\scrM,\Rmet)$, one can define a  Lorentzian geodesic space $\smash{(\scrM,\met^h,\ll_\Rmet^\leftarrow,\leq_\Rmet^\leftarrow,\tsep_\Rmet^\leftarrow)}$ in complete analogy to \autoref{Sec:Construction} relative to \emph{past-directed} --- in the evident sense --- in place of \emph{future-directed} $\Rmet$-timelike and $\Rmet$-causal curves. This is called the \emph{$\Rmet$-causally reversed} structure of $\smash{(\scrM,\met^h,\ll_\Rmet,\leq_\Rmet,\tsep_\Rmet)}$ \cite[Sec.~1.1]{cavalletti2020}. The regularity properties from \autoref{Pr:Properties} transfer to it.

Replacing $t$ by $1-t$ in \autoref{Def:TCD} and employing that this definition is ``symmetric'' in the regularity properties asked for $\mu_0$ and $\mu_1$, it is clear that $\smash{\scrX_\Rmet^V}$ satisfies $\smash{\TCD_p(K,N)}$ if and only if $\smash{(\scrX_\Rmet^V)^\leftarrow}$ does. A similar property for $\smash{\TMCP(K,N)}$ is unclear, for $\smash{\TMCP(K,N)}$ for $\smash{(\scrX_\Rmet^V)^\leftarrow}$ encodes semiconvexity of the Rényi entropy along timelike $\smash{\ell_{\Rmet,p}}$-optimal transport [sic] from a Dirac measure to an $\smash{\mmeas_{\Rmet}^V}$-absolutely continuous mass distribution.
\end{remark}

\section{Main results and consequences}\label{Ch:Proofs}

\subsection{Statements} Now we are in a position to state our main results in a more complete form than outlined in \autoref{Ch:Intro}.

\begin{theorem}\label{Th:TMCP} Assume $\smash{\Rmet}$ to be  $\smash{\Cont^1}$. Let $K\in\R$ and $N\in[n,\infty)$, and suppose  
\begin{align}\label{Eq:RNKrrr}
\Ric_\Rmet^{N,V} \geq K\text{ in all timelike directions}. 
\end{align}
Then the measured Lorentzian space $\smash{\scrX_\Rmet^V}$ from \eqref{Eq:X def} induced by $\smash{(\scrM,\Rmet,\mmeas_\Rmet^V)}$ satisfies $\smash{\TMCP(K,N)}$ according to \autoref{Def:TMCP}.

That is, for every $\smash{\mu_0 = \rho_0\,\mmeas_\Rmet^V\in \scrP_\comp^\ac(\scrM,\vol_\Rmet)}$ and every $x_1\in\scrM$ with $\smash{\mu_0[I_\Rmet^-(x_1)]}=1$, there are $p\in (0,1)$ and a timelike proper-time parametrized $\smash{\ell_{\Rmet,p}}$-geodesic $(\mu_t)_{t\in[0,1]}$ from $\mu_0$ to $\smash{\mu_1 := \delta_{x_1}}$ such that for every $t\in[0,1)$ and every $N'\geq N$,
\begin{align}\label{Eq:M1}
\scrS_\Rmet^{N',V}(\mu_t) \leq -\int_\scrM\tau_{K,N'}^{(1-t)}(\tsep_\Rmet(x^0,x_1))\,\rho_0(x^0)^{-1/N'}\d\mu_0(x^0).
\end{align}
\end{theorem}

\begin{theorem}\label{Th:TCD nonbr} Assume $\smash{\Rmet}$ to be $\smash{\Cont^{1,1}}$. Let $K\in\R$ and $N\in[n,\infty)$, and suppose \eqref{Eq:RNKrrr}. 
Then for every $p\in (0,1)$, the measured Lorentzian space $\smash{\scrX_\Rmet^V}$ from \eqref{Eq:X def} in\-duced by $\smash{(\scrM,\Rmet,\mmeas_\Rmet^V)}$ satisfies $\smash{\TCD_p(K,N)}$ according to \autoref{Def:TCD}.

That is, for all $\Rmet$-timelike $p$-dualizable  $\smash{(\mu_0,\mu_1) = (\rho_0\,\mmeas_\Rmet^V,\rho_1\,\mmeas_\Rmet^V)}\in \smash{\scrP_\comp^\ac(\scrM,\vol_\Rmet)^2}$, there exist
\begin{itemize}
\item a timelike proper-time parametrized $\smash{\ell_{\Rmet,p}}$-geodesic $(\mu_t)_{t\in[0,1]}$ connecting $\mu_0$ to $\mu_1$, and
\item a $\Rmet$-timelike $p$-dualizing coupling $\smash{\pi\in\Pi_{\ll_\Rmet}(\mu_0,\mu_1)}$
\end{itemize}
such that for every $t\in[0,1]$ and every $N'\geq N$,
\begin{align}\label{Eq:M2}
\begin{split}
\scrS_\Rmet^{N',V}(\mu_t) &\leq -\int_{\scrM^2}\tau_{K,N'}^{(1-t)}(\tsep_\Rmet(x^0,x^1))\,\rho_0(x^0)^{-1/N'}\d\pi(x^0,x^1)\\
&\qquad\qquad -\int_{\scrM^2} \tau_{K,N'}^{(t)}(\tsep_\Rmet(x^0,x^1))\,\rho_1(x^1)^{-1/N'}\d\pi(x^0,x^1).
\end{split}
\end{align}
\end{theorem}

\subsection{Proofs of \autoref{Th:TMCP} and \autoref{Th:TCD nonbr}} In order to prove the two preceding claims, the main work has to be performed for the more general version of \autoref{Th:TheMainTechnicalResult} stated now in \autoref{Pr:pr}, adapted to general $K$, $V$, and $N$.

For convenience, we will write
\begin{align*}
\Rmet_\infty := \Rmet.
\end{align*}
To relax notation a bit, we will agree that whenever a Lorentzian metric, say $\Rmet_k$, has a subscript $k\in \N_\infty$, we endow corresponding quantities defined by that metric with the same subscript, e.g.~we write $\vert\cdot\vert_k$ instead of $\smash{\vert\cdot\vert_{\Rmet_k}}$, etc.

\begin{proposition}\label{Pr:pr} Given any $K\in\R$ and $N\in [n,\infty)$, suppose 
\begin{align}\label{Eq:RIC BD}
\Ric_\infty^{N,V} \geq K\text{ in all timelike directions}.
\end{align}
Assume $\smash{(\mu_{\infty,0},\mu_{\infty,1}) = (\rho_{\infty,0}\,\mmeas_\infty^V,\rho_{\infty,1}\,\mmeas_\infty^V)\in \scrP_\comp^\ac(\scrM,\vol_\infty)^2}$ to satisfy 
\begin{align*}
\supp\mu_{\infty,0}\times\supp\mu_{\infty,1}\subset \scrM_{\ll_\infty}^2
\end{align*}
and $\rho_{\infty,0},\rho_{\infty,1}\in\Ell^\infty(\scrM,\vol_\infty)$. Then for every $p\in (0,1)$ there exist 
\begin{itemize}
\item a time\-like proper-time para\-met\-rized $\smash{\ell_{\infty,p}}$-geodesic $(\mu_{\infty,t})_{t\in [0,1]}$ from $\mu_{\infty,0}$ to $\mu_{\infty,1}$, and 
\item an $\smash{\ell_{\infty,p}}$-optimal coupling $\smash{\pi\in\Pi_{\ll_\infty}(\mu_{\infty,0},\mu_{\infty,1})}$  
\end{itemize}
such that for every $t\in[0,1]$ and every $N'\geq N$,
\begin{align*}
\scrS_{\infty}^{N',V}(\mu_{\infty,t}) &\leq -\int_{\scrM^2}\tau_{K,N'}^{(1-t)}(\tsep_\infty(x^0,x^1))\,\rho_0(x^0)^{-1/N'}\d\pi(x^0,x^1)\\
&\qquad\qquad -\int_{\scrM^2}\tau_{K,N'}^{(t)}(\tsep_\infty(x^0,x^1))\,\rho_1(x^1)^{-1/N'}\d\pi(x^0,x^1).
\end{align*}
\end{proposition}

The proof of this proposition, in turn, is subdivided into various subsections incorporated in the body of the text below.

To streamline the exposition, in this chapter we adopt the subsequent no\-tational convention. If a  quantity is not introduced explicitly in a specific result or proof, it automatically refers to the respective object defined in one of the results or proofs listed in this chapter. Also, until \autoref{Sec:Concl} various subsequences will be extracted, which is not notationally reflected either for readability.

\subsubsection{Setup and notation} Given the estimate \eqref{Eq:RIC BD}, let $(\varepsilon_k)_{k\in\N}$ be a fixed sequence in $(0,\infty)$ decreasing to $0$, let $\{\check{\Rmet}_\varepsilon : \varepsilon > 0\}$ be a family of smooth Lorentzian metrics satisfying all properties of \autoref{Le:Approx}, and set
\begin{align*}
\Rmet_k := \check{\Rmet}_{\varepsilon_k}.
\end{align*}
For $k\in\N_\infty$, according to \eqref{Eq:X def} we write
\begin{align*}
\scrX_k^V := (\scrM,\met^h,\mmeas_k^V,\ll_k,\leq_k,\tsep_k).
\end{align*}

In the sequel, we set
\begin{align}\label{Eq:kappa}
3\kappa := \inf\tsep_\infty(\supp\mu_{\infty,0}\times\supp\mu_{\infty,1})>0.
\end{align}

\subsubsection{Uniform convergence} In this technical section, we recapitulate the uniform convergence of $(\tsep_k)_{k\in\N}$ to $\tsep_\infty$ on compact subsets of $\smash{\scrM_{\ll_\infty}}$, cf.~\autoref{Cor:Unif cvg}. This will be needed in the proof of \autoref{Le:recovery sequence}, cf.~\eqref{Eq:Inclusion Omega}. For a similar result coming from approximation of the reference metric by smooth metrics with \emph{wider} light cones, see \cite[Prop.~A.2]{ms}.

\NEW{}{The proofs of the corresponding results are standard, hence omitted.}

\begin{lemma}\label{Le:tauinfty tauk} For every $\varepsilon > 0$ and every compact $\smash{C\subset\scrM_{\ll_\infty}^2}$, there exists $k_0\in\N$ such that for every $k\geq k_0$ and every $(x,y)\in C$,
\begin{align*}
\tsep_\infty(x,y) \leq \tsep_k(x,y) + \varepsilon.
\end{align*}
\end{lemma}


\begin{corollary}\label{Cor:Unif cvg} For every set $C$ as in \autoref{Le:tauinfty tauk}, the sequence $(\tsep_k)_{k\in\N}$ converges to $\tsep_\infty$ uniformly on $C$.
\end{corollary}


\subsubsection{Construction of a recovery sequence}\label{Sec:Recovery} Before getting to \autoref{Le:recovery sequence}, some further notational preparations are in order. 

Let $\mms$ be a $\smash{\met^h}$-closed ball in $\scrM$ which compactly contains $J_\infty(\mu_{\infty,0},\mu_{\infty,1})$. Since $\mmeas_\infty[\partial\mms] = 0$, by Portmanteau's theorem the sequence $(\meas_k)_{k\in\N}$ converges weakly to $\meas_\infty$, where we set, for $k\in\N_\infty$,
\begin{align*}
\meas_k := \mmeas_k^V[\mms]^{-1}\,\mmeas_k^V\mres M.
\end{align*}
Since $\mms$ is compact, $W_2(\meas_k,\meas_\infty)\to 0$ as $k\to\infty$, where $W_2$ is the $2$-Wasserstein metric on $\scrP(\mms)$ with respect to the restriction of $\smash{\met^h}$ to $\mms$. Given any $k\in\N$, let $\mathfrak{q}_k\in\scrP(\mms^2)$ be a fixed $\smash{W_2}$-optimal coupling of $\meas_k$ and $\meas_\infty$ \cite[Thm.~4.1]{villani2009}. Let $\smash{\mathfrak{p}^k\colon \mms\to\scrP(\mms)}$ denote the disintegration of $\mathfrak{q}_k$ with respect to $\pr_1$,  given by the formula $\rmd \mathfrak{q}_k(x,y) = \rmd\mathfrak{p}_x^k(y)\d\meas_k(x)$. 
Let $\smash{\mathfrak{p}^k \colon \scrP^\ac(\mms,\meas_\infty) \to \Prob^\ac(\mms,\meas_k)}$ denote the canonically induced (and nonrelabeled) map.

The proof of \autoref{Le:recovery sequence} below follows Step 1 to Step 3 for \cite[Thm.~3.29]{braun2022}. It involves the subsequent \autoref{Le:pin} \cite[Lem.~3.15]{cavalletti2020}. Various items listed therein do not explicitly appear in our arguments below, but are used in the out\-sourced parts of the proof of \autoref{Pr:pr} in \autoref{Sec:Concl}.

\begin{lemma}\label{Le:pin} Let $\pi_\infty\in\Pi_{\ll_\infty}(\mu_{\infty,0},\mu_{\infty,1})$ be $\Rmet_\infty$-timelike $p$-dualizing, $p\in (0,1]$. Then there exist sequences $(\pi_\infty^n)_{n\in\N}$ in $\scrP(\mms^2)$ and $(a_n)_{n\in\N}$ in $[1,\infty)$ such that
\begin{enumerate}[label=\textnormal{(\roman*)}]
\item the sequence $(a_n)_{n\in\N}$ converges to $1$,
\item $\smash{\pi_\infty^n[\mms_{\ll_\infty}^2] = 1}$ for every $n\in\N$,
\item $\smash{\pi_\infty^n = \rho_\infty^n\,\meas_\infty^{\otimes 2}\in\Prob^\ac(\mms^2,\meas_\infty^{\otimes 2})}$ and $\smash{\rho_\infty^n\in\Ell^\infty(\mms^2,\meas_\infty^{\otimes 2})}$ for every $n\in\N$,
\item the sequence $(\pi_\infty^n)_{n\in\N}$ converges weakly to $\pi_\infty$,
\item writing $\smash{\rho_{\infty,0}^n}$ and $\smash{\rho_{\infty,1}^n}$ for the density of the first and second marginal of $\smash{\pi_\infty^n}$ with respect to $\meas_\infty$, we have 
\begin{alignat*}{3}
\rho_{\infty,0}^n &\leq a_n\,\rho_{\infty,0} &\quad &\meas_\infty\text{-a.e.},\\
\rho_{\infty,1}^n &\leq a_n\,\rho_{\infty,1}  & &\meas_\infty\text{-a.e.},
\end{alignat*}
\item $\smash{\rho_{\infty,0}^n\to \rho_{\infty,0}}$ and $\rho_{\infty,1}^n \to \rho_{\infty,1}$ in $\Ell^1(\mms,\meas_\infty)$ as $n\to\infty$.
\end{enumerate}
\end{lemma}

\begin{lemma}\label{Le:recovery sequence} Let $p\in (0,1]$. Then there exists a sequence $(\mu_{k,0},\mu_{k,1})_{k\in\N}$ of pairs $\smash{(\mu_{k,0},\mu_{k,1}) = (\rho_{k,0}\,\meas_k,\rho_{k,1}\,\meas_k)\in\scrP^\ac(\mms,\meas_k)}$ such that
\begin{enumerate}[label=\textnormal{(\roman*)}]
\item $(\mu_{k,0},\mu_{k,1})_{k\in\N}$ converges weakly to $(\mu_{\infty,0},\mu_{\infty,1})$, and
\item for every $k\in\N$, the pair $(\mu_{k,0},\mu_{k,1})$ is $\Rmet_k$-timelike $p$-dualizable by a coupling $\bar{\pi}_k\in\smash{\Pi_{\ll_k}(\mu_{k,0},\mu_{k,1})}$ satisfying 
\begin{align*}
\bar{\pi}_k[\{\tsep_k > \kappa\}]=1.
\end{align*}
\end{enumerate}
\end{lemma}

\begin{proof} Given a $\Rmet$-timelike $p$-dualizing coupling $\pi_\infty\in\smash{\Pi_{\ll_\infty}(\mu_{\infty,0},\mu_{\infty,1})}$, let $(\pi^n_\infty)_{n\in\N}$ be as in \autoref{Le:pin}. Define $\smash{\mu_{k,0}^n,\mu_{k,1}^n\in\Prob^\ac(\mms,\meas_k)}$, $k\in\N$, by
\begin{align*}
\mu_{k,0}^n &:= \mathfrak{p}^k(\mu_{\infty,0}^n) = \rho_{k,0}^n\,\meas_k,\\
\mu_{k,1}^n &:= \mathfrak{p}^k(\mu_{\infty,1}^n) = \rho_{k,1}^n\,\meas_k.
\end{align*}
Moreover, define $\smash{\pi_k^n\in\Pi(\mu_{k,0}^n,\mu_{k,1}^n) \cap \Prob^\ac(\mms^2,\meas_k^{\otimes 2})}$ by
\begin{align*}
\pi_k^n := (\pr_1,\pr_3)_\push\big[(\rho_\infty^n\circ(\pr_2,\pr_4))\,\mathfrak{q}_k\otimes\mathfrak{q}_k\big].
\end{align*}
Using tightness of $(\mathfrak{q}_k)_{k\in\N}$ \cite[Lem.~4.3, Lem.~4.4]{villani2009}, we obtain the weak convergence of $\smash{(\pi_k^n)_{k\in\N}}$ to $\pi_\infty^n$, $n\in\N$, up to a nonrelabeled subsequence. Then \autoref{Le:pin}, a compactness argument, and a diagonal procedure yield a sequence $(\tilde{\pi}_k)_{k\in\N}$ of probability measures $\smash{\tilde{\pi}_k\in\Prob^\ac(\mms^2,\meas_k^{\otimes 2})}$ converging weakly to $\pi_\infty$, with
\begin{align*}
\tilde{\pi}_k := \pi_k^{n_k}.
\end{align*}

Let $U_0,U_1\subset \mms$ be relatively compact open sets with $\supp\mu_{\infty,0}\subset U_0$, $\supp\mu_{\infty,1}\subset U_1$, and $\inf\tsep_\infty(\bar{\Omega}) > 2\kappa$, where
\begin{align*}
\Omega := U_0 \times U_1.
\end{align*}
By \autoref{Le:tauinfty tauk} applied to $\varepsilon := \kappa$ and $\smash{C := \bar{\Omega}}$, we have 
\begin{align}\label{Eq:Inclusion Omega}
\Omega \subset \{\tsep_k > \kappa\}
\end{align}
for large enough $k\in\N$. By Portmanteau's theorem, since $\Omega$ is open,
\begin{align*}
1 = \pi_\infty[\Omega] \leq \liminf_{k\to\infty} \tilde{\pi}_k[\Omega].
\end{align*}
Up to passage to a subsequence, we may and will thus assume $\smash{\tilde{\pi}_k[\Omega] > 0}$ for every $k\in\N$. Let the marginals $\smash{\tilde{\mu}_{k,0},\tilde{\mu}_{k,1}\in\Prob^\ac(\mms,\meas_k)}$ of $\smash{\tilde{\pi}_k}$ be given by
\begin{align*}
\tilde{\mu}_{k,0} &= \tilde{\rho}_{k,0}\,\meas_k = \rho_{k,0}^{n_k}\,\meas_k,\\
\tilde{\mu}_{k,1} &= \tilde{\rho}_{k,1}\,\meas_k = \rho_{k,1}^{n_k}\,\meas_k.
\end{align*}

Define $\smash{\hat{\pi}_k \in\Prob^\ac(\mms,\meas_k)}$ through
\begin{align*}
\hat{\pi}_k := \tilde{\pi}_k[\Omega]^{-1}\,\tilde{\pi}_k\mres \Omega
\end{align*}
with marginals $\smash{\hat{\mu}_{k,0},\hat{\mu}_{k,1}\in \scrP^\ac(\mms,\meas_k)}$ given by
\begin{align*}
\hat{\mu}_{k,0} &= \hat{\rho}_{k,0}\,\meas_k,\\
\hat{\mu}_{k,1} &= \hat{\rho}_{k,1}\,\meas_k.
\end{align*}
Albeit these measures admit a $\Rmet_k$-chronological coupling by construction, it is not clear whether these are $\Rmet_k$-timelike $p$-dualizable, i.e.~their $\smash{\smash{\ell}_{k,p}}$-cost is maximized by a coupling concentrated on the set $\smash{\mms_{\ll_k}^2}$. To modify $\smash{\hat{\mu}_{k,0}}$ and $\smash{\hat{\mu}_{k,1}}$ accordingly, let $\check{\pi}_k\in\Pi_\leq(\hat{\mu}_{k,0},\hat{\mu}_{k,1})$ be an $\smash{\ell_{k,p}}$-optimal coupling; by choosing the previous coupling $\hat{\pi}_k$ as a competitor, and using  compactness of $\mms^2$, its cost is strictly positive and finite. Since $(\hat{\pi}_k)_{k\in\N}$ is weakly convergent, its marginal sequences $\smash{(\hat{\mu}_{k,0})_{k\in\N}}$ and $\smash{(\hat{\mu}_{k,1})_{k\in\N}}$ are tight; so is $(\check{\pi}_k)_{k\in\N}$ by \cite[Lem.~4.4]{villani2009}. Thus, a nonrelabeled subsequence of the latter converges weakly to some $\smash{\check{\pi}_\infty\in\Pi(\mu_{\infty,0},\mu_{\infty,1})}$. 
By \eqref{Eq:kappa},
\begin{align*}
1 = \check{\pi}_\infty[\Omega] \leq \liminf_{k\to\infty}\check{\pi}_k[\Omega].
\end{align*}
Up to passing to a subsequence, we may and will thus assume that $\smash{\check{\pi}_k[\Omega]>0}$ for every $k\in\N$. Then we define $\bar{\pi}_k \in\Prob(\mms^2)$ through
\begin{align*}
\bar{\pi}_k := \check{\pi}_k[\Omega]^{-1}\,\check{\pi}_k \mres \Omega.
\end{align*}
By the restriction property of $\smash{\ell_{k,p}}$-optimal couplings \cite[Lem.~2.10]{cavalletti2020}, $\smash{\bar{\pi}_k}$ constitutes a chronological $\smash{\ell_{k,p}}$-optimal coupling of its marginals $\mu_{k,0},\mu_{k,1}\in\Prob^\ac(\mms,\meas_k)$; in fact, $\bar{\pi}_k$ will even be uniquely determined by that property, see e.g.~the proof of \autoref{Pr:Displacement}. Moreover, we have $\smash{\bar{\pi}_k[\{\tsep_k>\kappa\}]=1}$ for large enough $k\in\N$ thanks to \eqref{Eq:Inclusion Omega}. Hence, the pair $(\mu_{k,0},\mu_{k,1})$ and $\bar{\pi}_k$ obey  the desired requirements.
\end{proof}

\subsubsection{Displacement semiconvexity}\label{Sec:Displacement} Now we prove displacement semiconvexity of Rényi's entropy with respect to $\meas_k$ between $\mu_{k,0}$ and $\mu_{k,1}$. In view of \autoref{Le:Approx}, this is the point where the additional property $\smash{\bar{\pi}_k[\{\tsep_k > \kappa\}]=1}$ for every $k\in\N$, independently of the value $\kappa$ from \eqref{Eq:kappa}, from \autoref{Le:recovery sequence}  comes into play.

In the sequel, let $\smash{\scrS_k^N}$ denote the $N$-Rényi entropy with respect to $\meas_k$, $k\in\N_\infty$, defined analogously to \eqref{Eq:Renyi}.

\begin{proposition}\label{Pr:Displacement} Let $\delta > 0$. Then there exists $k_0\in\N$ such that for every  $k\in\N$ with $k\geq k_0$, there exists a timelike proper-time parametrized $\smash{\ell_{k,p}}$-geodesic $(\mu_{k,t})_{t\in[0,1]}$ from $\mu_{k,0}$ to $\mu_{k,1}$ such that for every $t\in[0,1]$ and every $N'\geq N$,
\begin{align}\label{Eq:THEST}
\begin{split}
\scrS_k^{N'}(\mu_{k,t}) &\leq -\int_{\mms^2}\tau_{K-\delta,N'}^{(1-t)}(\tsep_k(x^0,x^1))\,\rho_{k,0}(x^0)^{-1/N'}\d\bar{\pi}_k(x^0,x^1)\\
&\qquad\qquad -\int_{\mms^2} \tau_{K-\delta,N'}^{(t)}(\tsep_k(x^0,x^1))\,\rho_{k,1}(x^1)^{-1/N'}\d\bar{\pi}_k(x^0,x^1).
\end{split}
\end{align}
\end{proposition}

\begin{proof} The claim follows from essentially the same computations as \cite[Ch.~6]{mccann2020} \NEW{}{and \cite[Thm.~5.9]{braunohta}}. We only describe the setting and the necessary modifications.

Let $c>0$ be a given constant with respect to which all $\Rmet_\infty$-causal curves passing through the compact set $\mms$ have $\smash{\met^h}$-length no larger than $c$. (Thus, all $\Rmet_k$-causal curves with endpoints in $\mms$ do not leave that set by \autoref{Le:Approx}, $k\in\N$, which will be used several times without explicit notice below.) For such $c$, $\delta$ as hypothesized, $\mms$ as given, and $\kappa$ as in \eqref{Eq:kappa}, let $k_0\in\N$ be as provided by \autoref{Le:Approx}. Let $k\in\N$ with $k\geq k_0$, and recall from \autoref{Le:Glob hyperb} that $\Rmet_k$ is globally hyperbolic. Hence, the theory developed in \cite{mccann2020} applies as follows. As $\smash{\bar{\pi}_k}$ is chronological and $\smash{\ell_{k,p}}$-optimal, standard Kantorovich duality, cf.~\cite[Thm.~5.10]{villani2009} and \cite[Rem.~2.2, Prop.~2.8, Prop.~2.19]{cavalletti2020}, entails the $p$-separation of $(\mu_{k,0},\mu_{k,1})$ according to \cite[Def.~4.1]{mccann2020}. Since $\mu_{k,0} \ll \meas_k \ll \vol_k$, $\bar{\pi}_k$ is the \emph{unique} chronological $\smash{\ell_{k,p}}$-optimal coupling of $\mu_{k,0}$ and $\mu_{k,1}$ relative to the Lorentzian spacetime $(\scrM,\Rmet_k)$ \cite[Thm.~5.8]{mccann2020}. In particular, there is a sufficiently regular vector field $X_k$ on $\scrM$ such that
\begin{align*}
\bar{\pi}_k = (\Id,T_{k,1})_\push\mu_{k,0},
\end{align*}
where $\smash{T_{k,\cdot}\colon[0,1]\times\mms\to\mms}$ is given by
\begin{align*}
T_{k,t}(x) := \exp_x t X_k(x).
\end{align*}
Moreover, by \cite[Rem.~B.10]{braun2022} and \cite[Cor.~5.9]{mccann2020}, there exists a unique timelike proper-time parametrized $\smash{\ell_{k,p}}$-geodesic $(\mu_{k,t})_{t\in[0,1]}$ from $\mu_{k,0}$ to $\mu_{k,1}$. It is given by
\begin{align}\label{Eq:mukt}
\mu_{k,t} = (T_{k,t})_\push\mu_{k,0}.
\end{align}

Lastly, let $\smash{A_{k,t} := \tilde{\rmD}T_{k,t}\colon T\scrM\big\vert_\mms \to (T_{k,t})_*T\scrM}$ be the approximate derivative \cite[Def. 3.8]{mccann2020} of $T_{k,t}$ as given by \cite[Prop.~6.1]{mccann2020}. It is invertible and depends smoothly on $t\in[0,1]$ at $\vol_k$-a.e.~$x\in\mms$. For such $x$ and a given $t\in[0,1]$, set
\begin{align*}
\jmath_{k,t}(x) &:= \vert\!\det A_{k,t}(x)\vert\,\rme^{-V(T_{k,t}(x))},\\
\varphi_{k,t}(x) &:= \log\jmath_{k,t}(x) = \log\vert\!\det A_{k,t}(x)\vert - V(T_{k,t}(x)).
\end{align*}

Assume $N' \geq N > n$; the case $N=n$ can be treated similary. Evaluated at any fixed point in $\mms$, the curve $(T_{k,t})_{t\in[0,1]}$ is a $\Rmet_k$-timelike geodesic passing through $\mms$. In particular, its $\smash{\met^h}$-length is no larger than $c$, whence $\smash{\vert\dot{T}_{k,t}\vert_h \leq c}$ for every given $t\in[0,1]$. Moreover, geodesy \cite[Thm.~6.4]{mccann2020}, \eqref{Eq:mukt}, and $\smash{\bar{\pi}_k[\{\tsep_k > \kappa\}] = 1}$ imply
\begin{align}\label{Eq:thetak}
\vartheta_k := \vert\dot{T}_{k,t}\vert_k = \tsep_k(\cdot,T_{k,1}) > \kappa\quad\mu_{k,0}\text{-a.e.}
\end{align}
Computing as in Step 2 for \cite[Prop.~A.2]{braun2022} and using \autoref{Le:Approx} with \eqref{Eq:thetak}, 
\begin{align*}
\ddot{\varphi}_{k,t} + \frac{1}{N'}\,\dot{\varphi}_{k,t}^2 &\leq \ddot{\varphi}_{k,t} + \frac{1}{N}\,\dot{\varphi}_{k,t}^2 \\
&\leq -\Ric_k^{N,V}(\dot{T}_{k,t},\dot{T}_{k,t}) \leq -(K-\delta)\,\vartheta_k^2\quad\mu_{k,0}\text{-a.e.}
\end{align*}
This is a version of (A.4) in \cite{braun2022}. From here, we follow the proof of \cite[Thm.~5.9]{braunohta} \emph{verbatim}  
to conclude the statement.
\end{proof}

\subsubsection{Conclusions}\label{Sec:Concl} For notational convenience, given any $\pi\in\Pi(\mu_{\infty,0},\mu_{\infty,1})$, $t\in[0,1]$, $K\in\R$, and $N\in[1,\infty)$, we define
\begin{align*}
\scrT_{K,N}^{(t)}(\pi) &:= -\int_{\mms^2} \tau_{K,N}^{(1-t)}(\tsep_\infty(x^0,x^1))\,\rho_{\infty,0}(x^0)^{-1/N'}\d\pi(x^0,x^1)\\
&\qquad\qquad -\int_{\mms^2}\tau_{K,N}^{(1-t)}(\tsep_\infty(x^0,x^1))\,\rho_{\infty,1}(x^1)^{-1/N'}\d\pi(x^0,x^1).
\end{align*}

\begin{proof}[Proof of \autoref{Pr:pr}] The estimate obtained in \autoref{Pr:Displacement} is a version of (3.9) in \cite{braun2022}, with $\smash{\pi_k := \bar{\pi}_k}$, $k\in\N$ with $k\geq k_0$. Given any $\eta > 0$, by \autoref{Cor:Unif cvg} we can modify $k_0$ in such a way that for every $k\geq k_0$, $\tsep_k$ can be replaced by $\tsep_\infty$ on the right-hand side of \eqref{Eq:THEST}, and the two respective expressions differ at most by $\eta$. From there, 
letting $k\to\infty$ for \emph{fixed} $\delta,\eta > 0$ we follow \emph{verbatim} the proof of \cite[Thm.~3.29]{braun2022} --- with $\tsep_\infty$ in place of $\tsep$ therein --- and get the following property. Given $\delta$ and $\eta$ as above, there exist a time\-like proper-time parametrized $\smash{\ell_{\infty,p}}$-geodesic $\smash{(\mu_{\infty,t}^\delta)_{t\in[0,1]}}$ from $\mu_{\infty,0}$ to $\mu_{\infty,1}$ and a $\smash{\Rmet_\infty}$-timelike $p$-dualizing coupling $\smash{\pi_\infty^\delta \in \Pi_{\ll_\infty}(\mu_{\infty,0},\mu_{\infty,1})}$ such that for every $t\in[0,1]$ and every $N'\geq N$, we have
\begin{align}\label{Eq:Estimate}
\scrS_\infty^{N'}(\mu_{\infty,t}^\delta) \leq\scrT_{K-\delta,N'}^{(t)}(\pi_\infty^\delta) + \eta.
\end{align}
Note that the inherent objects do not depend on $\eta$.

Fix a sequence $(\delta_n)_{n\in\N}$ in $(0,\infty)$  decreasing to $0$, and let $\smash{(\mu_{\infty,t}^{\delta_n})_{t\in[0,1]}}$ and $\smash{\pi_\infty^{\delta_n}}$ be the above objects with respect to $\delta_n$, $n\in\N$. Let $\bdpi^n\in\smash{\OptTGeo_{\ell_{\infty,p}}^{\tsep_\infty}(\mu_{\infty,0},\mu_{\infty,1})}$ represent $\smash{(\mu_{\infty,t}^{\delta_n})_{t\in[0,1]}}$. By our assumption 
\begin{align*}
\supp\mu_{\infty,0}\times\supp\mu_{\infty,1}\subset\mms_{\ll_\infty}^2
\end{align*}
and by compactness of timelike $\smash{\ell_{\infty,p}}$-optimal geodesic plans relative to $\smash{\scrX_\infty^V}$ constructed in \autoref{Sec:Construction} \cite[Prop.~B.11]{braun2022}, cf.~\autoref{Pr:Properties}, a nonrelabeled subsequence of $\smash{(\bdpi^n)_{n\in\N}}$ converges weakly to some  $\smash{\bdpi\in\OptTGeo_{\ell_{\infty,p}}^{\tsep_\infty}(\mu_{\infty,0},\mu_{\infty,1})}$. The latter represents a timelike proper-time pa\-ra\-metrized $\smash{\ell_{\infty,p}}$-geodesic $(\mu_{\infty,t})_{t\in[0,1]}$ from $\mu_{\infty,0}$ to $\mu_{\infty,1}$. Moreover, by a tightness argument and stability of $\smash{\ell_{\infty,p}}$-optimal couplings \cite[Lem.~2.11]{cavalletti2020}, a nonrelabeled subsequence of $\smash{(\pi_\infty^{\delta_n})_{n\in\N}}$ converges weakly to some $\smash{\ell_{\infty,p}}$-optimal coupling $\pi_\infty\in\Pi_\ll(\mu_{\infty,0},\mu_{\infty,1})$. Thus, given $\varepsilon > 0$, $t\in[0,1]$, and $N'\geq N$ we obtain
\begin{align*}
\scrS_\infty^{N'}(\mu_{\infty,t}) &\leq \limsup_{n\to\infty}\scrS_\infty^{N'}(\mu_{\infty,t}^{\delta_n}) \leq \limsup_{n\to\infty}\scrT_{K-\delta_n,N'}^{(t)}(\pi_\infty^{\delta_n}) +\eta\\
&\leq \limsup_{n\to\infty} \scrT_{K-\varepsilon,N'}^{(t)}(\pi_\infty^{\delta_n}) +\eta \leq \scrT_{K-\varepsilon,N'}^{(t)}(\pi_\infty) + \eta.
\end{align*}
Here we have successively used weak lower semicontinuity of the Rényi entropy on $\scrP(\mms)$ \cite[Thm.~B.33]{lott2009}, the estimate \eqref{Eq:Estimate}, nondecreasingness of the distortion coefficient $\smash{\tau_{K,N'}^{(r)}(\vartheta)}$ in $K\in\R$ for fixed $r\in [0,1]$, $N' \geq N$, and $\vartheta\geq 0$, as well as upper semicontinuity of $\smash{\scrT_{K-\varepsilon,N'}^{(t)}}$ after \cite[Lem.~3.27]{braun2022}. Finally, sending $\varepsilon\to 0$ and $\eta\to 0$ in the previous inequality via Fatou's lemma gives the result.
\end{proof}

\begin{proof}[Proof of \autoref{Th:TMCP}] Combining \autoref{Pr:pr} with \cite[Prop.~4.9]{braun2022}, we directly  obtain the $\smash{\TMCP(K,N)}$ condition for $\smash{\scrX_\infty^V}$. Indeed, albeit \cite[Prop.~4.9]{braun2022} assumes the \emph{weak} timelike curvature-dimension condition from \cite[Def.~3.3]{braun2022}, its proof needs displacement semiconvexity of the Rényi entropy only between mass distributions satisfying the assumptions of \autoref{Pr:pr}.
\end{proof}

\begin{proof}[Proof of \autoref{Th:TCD nonbr}] Recall from \autoref{Re:Geod equ} that if $\Rmet_\infty$ is of class $\smash{\Cont^{1,1}}$, then $\smash{\scrX_\infty^V}$ is $\Rmet_\infty$-timelike nonbranching. Up to a change of the involved distortion coefficients, the identical argument as for \cite[Prop.~3.38]{braun2022} --- note that the reductions in Step 1 therein are precisely the assumptions on the marginals in \autoref{Pr:pr} --- entails a pathwise version of $\smash{\TCD_p(K,N)}$. This verifies $\smash{\TCD_p(K,N)}$ by integration.
\end{proof}

\subsection{Consequences of \autoref{Th:TMCP} and \autoref{Th:TCD nonbr}} 

\subsubsection{Sharp timelike geometric inequalities} Having established displacement semiconvexity of $\smash{\scrS_\Rmet^{N,V}}$ along appropriate timelike proper-time parametrized $\smash{\ell_{\Rmet,p}}$-geodesics, the following three geometric inequalities are derived in a standard way, cf.~\cite[Prop. 3.4, Prop.~3.5, Prop.~3.6]{cavalletti2020} or \cite[Prop.~3.11, Cor.~3.14, Thm.~3.16]{braun2022}. For instance, \autoref{Cor:Sharp1} is a simple consequence of  Jensen's inequality.

\begin{corollary}[Sharp Brunn--Minkowski]\label{Cor:Sharp1} Let the assumptions of \autoref{Th:TMCP} hold.  Let $p\in (0,1)$, let $A_0 \subset\scrM$ be a relatively compact Borel set with $\smash{\mmeas_\Rmet^V[A_0]>0}$, and let $\mu_0\in\scrP_\comp^\ac(\scrM,\vol_\Rmet)$ be the uniform distribution on $A_0$. For a specified Borel set $A_1\subset\scrM$ and  $t\in[0,1]$, we set
\begin{align*}
A_t := \{\gamma_t : \gamma\in\TGeo^{\tsep_\Rmet}(\scrM),\, \gamma_0\in A_0,\, \gamma_1\in A_1\}
\end{align*}
as well as
\begin{align*}
\Theta := \begin{cases}  \sup\tsep_\Rmet(A_0\times A_1) & \text{if }K<0,\\
\inf\tsep_\Rmet(A_0\times A_1) &\textnormal{otherwise}.
\end{cases}
\end{align*}
\begin{enumerate}[label=\textnormal{(\roman*)}]
\item Let $x_1\in\scrM$ such that $\mu_0[I_\Rmet^-(x_1)]=1$, and set $A_1 := \{x_1\}$. Then for every $t\in [0,1)$ and every $N'\geq N$,
\begin{align*}
\mmeas_\Rmet^V[A_t]^{1/N'} \geq \tau_{K,N'}^{(1-t)}(\Theta)\,\mmeas_\Rmet^V[A_0]^{1/N'}.
\end{align*}
\item Let the assumptions of \autoref{Th:TCD nonbr} hold. Let $A_1\subset\scrM$ be a relatively compact Borel set with $\smash{\mmeas_\Rmet^V[A_1] >0}$. Let $\mu_1\in\scrP_\comp^\ac(\scrM,\vol_\Rmet)$ be the uniform distribution on $A_1$, and assume $\Rmet$-timelike $p$-dualizability of $(\mu_0,\mu_1)$. Then for every $t\in[0,1]$ and every $N'\geq N$,
\begin{align*}
\mmeas_\Rmet^V[A_t]^{1/N'} \geq  \tau_{K,N'}^{(1-t)}(\Theta)\,\mmeas_\Rmet^V[A_0]^{1/N'} + \tau_{K,N'}^{(t)}(\Theta)\,\mmeas_\Rmet^V[A_1]^{1/N'}.
\end{align*}
\end{enumerate}
\end{corollary}

For a general $\mu_0$ as above, there might be no  $x_1\in\scrM$ with $\smash{\mu_0[I_\Rmet^-(x_1)]=1}$, thus its existence in the first item of \autoref{Cor:Sharp1} is an additional assumption. 

\begin{corollary}[Sharp Bonnet--Myers]\label{Cor:Sharp2} Let the assumptions from \autoref{Th:TMCP} hold, and further suppose $K>0$. Then
\begin{align*}
\sup\tsep_\Rmet(\scrM^2) \leq \pi\sqrt{\frac{N-1}{K}}.
\end{align*}
\end{corollary}

For the third corollary, we refer to \eqref{Eq:Distortion coeff} for the definition of the function $\smash{\mathfrak{s}_{K,N}}$.  Moreover, we call a set $E\subset \scrM$ $\tsep_\Rmet$-star-shaped with respect to $x\in\scrM$ if for every $\smash{\gamma\in\TGeo^{\tsep_\Rmet}(\scrM)}$ with $\gamma_0=x$ and $\gamma_1\in E$ we have $\gamma_t\in E$ for every $t\in(0,1)$. Given such $E$ and $x$ as well as $r>0$, set 
\begin{align*}
\sfB^{\tsep_\Rmet}(x,r) := \{y\in\scrM : \tsep_\Rmet(x,y) \in (0,r)\} \cup\{x\},
\end{align*}
and define 
\begin{align*}
v_r &:=\mmeas_\Rmet^V\big[\bar{\sfB}^{\tsep_\Rmet}(x,r)\cap E\big],\\
s_r &:=\limsup_{\delta\rightarrow0}\delta^{-1}\,\mmeas_\Rmet^V\big[(\bar{\sfB}^{\tsep_\Rmet}(x,r+\delta)\setminus\sfB^{\tsep_\Rmet}(x,r))\cap E\big].
\end{align*}

\begin{corollary}[Sharp Bishop--Gromov]\label{Cor:Sharp3} Let the assumptions of \autoref{Th:TMCP} hold. Let $E\subset \scrM$ be a compact set which is $\tsep_\Rmet$-star-shaped with respect
to $x\in\scrM$. Then for every $r,R>0$ with $\smash{r<R\leq\pi\sqrt{(N-1)/\max\{K,0\}}}$,
\begin{align*}
\frac{s_r}{s_R}\geq\Big[\frac{\mathfrak{s}_{K,N-1}(r)}{\mathfrak{s}_{K,N-1}(R)}\Big]^{N-1}
\end{align*}
as well as
\begin{align*}
\frac{v_r}{v_R}\geq\frac{\displaystyle\int_0^r\mathfrak{s}_{K,N-1}(s)^{N-1}\d s}{\displaystyle\int_0^R\mathfrak{s}_{K,N-1}(s)^{N-1}\d s}.
\end{align*}
\end{corollary}

\begin{remark}\label{Re:Sharp} These three corollaries explain why we chose to derive $\smash{\TMCP(K,N)}$ and $\smash{\TCD_p(K,N)}$ in \autoref{Th:TMCP} and \autoref{Th:TCD nonbr} instead of their reduced or entropic versions, cf.~\autoref{Re:RedEnt}. Indeed, \autoref{Cor:Sharp3} is sharp in the sense that model spaces attain equality therein \cite[Rem.~5.11]{cavalletti2020}. More generally, \autoref{Cor:Sharp1}, \autoref{Cor:Sharp2}, and \autoref{Cor:Sharp3} are sharp in the sense of dimensional improvements: recall that if a globally hyperbolic $\smash{\Cont^1}$-spacetime of dimension $n$ obeys $\smash{\TCD_p(K,N)}$ for some $N\in [1,\infty)$, then
\begin{align*}
n = \dim^{\tsep_\Rmet}\,\scrM \leq N.
\end{align*}
Here $\smash{\dim^{\tsep_\Rmet}\,\scrM}$ is the Lorentzian Hausdorff dimension of $(\scrM,\Rmet)$ from \cite[Def.~3.1]{ms}. Under $\smash{\TCD_p^*(K,N)}$ or $\smash{\TCD_p^e(K,N)}$, the above statements do \emph{a priori} only hold for $N$ replaced by $N+1$, cf.~\cite[Rem.~3.19]{braun2022} and \cite[Thm.~5.2]{ms}.

Under the stronger assumptions of \autoref{Th:TCD nonbr}, \autoref{Cor:Sharp2} and \autoref{Cor:Sharp3} follow alternatively from \autoref{Re:RedEnt} and \cite[Prop.~5.9, Prop.~5.10]{cavalletti2020}. The latter have been derived by using the localization technique from \cite[Ch.~4]{cavalletti2020}, itself reliant on $\Rmet$-timelike nonbranching, which may fail below $\smash{\Cont^{1,1}}$-regularity under synthetic timelike Ricci bounds \cite{gh}.
\end{remark}

\subsubsection{Uniqueness of chronological optimal couplings and chronological geodesics} A further direct implication of \autoref{Th:TCD nonbr}, together with the implicit $\Rmet$-timelike nonbranching property, are the following uniqueness results about the $\smash{\ell_{\Rmet,p}}$-optimal transport problem, cf.~\cite[Thm.~3.19, Thm.~3.20]{cavalletti2020} or \cite[Thm.~4.16, Thm.~4.17]{braun2022}.

\begin{corollary}\label{Cor:UNIQNNN} Let the assumptions of \autoref{Th:TCD nonbr} hold. Given $p\in (0,1)$, suppose $\Rmet$-timelike $p$-dualizability of the pair $\smash{(\mu_0,\mu_1)\in \scrP_\comp^\ac(\scrM,\vol_\Rmet) \times \scrP_\comp(\scrM)}$.
\begin{enumerate}[label=\textnormal{(\roman*)}]
\item \textnormal{\textbf{Uniqueness of chronological optimal couplings.}} The set of $\smash{\ell_{\Rmet,p}}$-op\-ti\-mal couplings of $\mu_0$ and $\mu_1$ which also lie in $\smash{\Pi_{\ll_\Rmet}(\mu_0,\mu_1)}$ is a singleton $\{\pi\}$. Moreover, there exists a $\mu_0$-measurable map $T\colon \supp\mu_0\to\scrM$ such that
\begin{align*}
\pi = (\Id,T)_\push\mu_0.
\end{align*}
\item \textnormal{\textbf{Uniqueness of chronological geodesics.}} The set $\smash{\OptGeo_{\ell_{\Rmet,p}}^{\tsep_\Rmet}(\mu_0,\mu_1)}$ is a singleton $\{\bdpi\}$. Furthermore, there exists a $\mu_0$-measurable map $\mathfrak{T}\colon \supp\mu_0 \to \smash{\TGeo^{\tsep_\Rmet}(\scrM)}$ such that
\begin{align*}
\bdpi = \mathfrak{T}_\push\mu_0.
\end{align*}
\end{enumerate}
\end{corollary}

\end{document}